\documentclass[english,twocolumn,twoside]{IEEEtran}
\usepackage[T1]{fontenc}
\usepackage[latin9]{inputenc}
\usepackage{color}
\usepackage{units}
\usepackage{mathtools}
\usepackage{amsthm}
\usepackage{amsmath}
\usepackage{amssymb}
\usepackage{graphicx}
\usepackage{esint}
\PassOptionsToPackage{normalem}{ulem}
\usepackage{ulem}
\usepackage{epstopdf}
\usepackage{tikz}
\usepackage{color}
\usetikzlibrary{patterns}
\usetikzlibrary{positioning}
\usepackage{csquotes}
\usepackage{todonotes}
\usepackage[normalem]{ulem} 
\presetkeys%
    {todonotes}%
    {inline}{}
\makeatletter
\usepackage[
style=ieee,
doi=false,isbn=false,url=false,sorting=nty
]{biblatex}

\theoremstyle{plain}
\newtheorem{thm}{\protect\theoremname}
\theoremstyle{plain}
\newtheorem{lem}[thm]{\protect\lemmaname}
\theoremstyle{definition}
\newtheorem{rem}[thm]{\protect\remarkname}

\usepackage{thmtools,url,psfrag,pstool}
\declaretheoremstyle[
spaceabove=3pt, spacebelow=3pt,
headfont=\normalfont\bfseries,
notefont=\mdseries, notebraces={(}{)},
bodyfont=\normalfont,
headformat=(\NAME\NUMBER),
postheadspace=.5em,
headpunct={}
]{mystyle}

\declaretheorem[style=mystyle,name=A,Refname=A,refname=A]{assumption}

\usepackage{mleftright}\mleftright
\date{}

\makeatother

\usepackage{babel}
\providecommand{\lemmaname}{Lemma}
\providecommand{\theoremname}{Theorem}
\providecommand{\remarkname}{Remark}
\newcommand{\eql}[2]{I(#1,\allowbreak#2)}

\newcommand{\cA}{\mathcal{A}}
\newcommand{\cB}{\mathcal{B}}

\newcommand{\cF}{\mathcal{F}}

\newcommand{\cH}{\mathcal{H}}

\newcommand{\cP}{\mathcal{P}}

\newcommand{\bbB}{\mathbb{B}}

\newcommand{\bbE}{\mathbb{E}}

\newcommand{\bbN}{\mathbb{N}}

\newcommand{\bh}{{\boldsymbol{h}}}

\newcommand{\bs}{{\boldsymbol{s}}}

\newcommand{\bu}{{\boldsymbol{u}}}
\newcommand{\bv}{{\boldsymbol{v}}}
\newcommand{\bw}{{\boldsymbol{w}}}
\newcommand{\bx}{{\boldsymbol{x}}}
\newcommand{\by}{{\boldsymbol{y}}}
\newcommand{\bz}{{\boldsymbol{z}}}

\newcommand{\bdelta}	{\boldsymbol{\delta}}

\newcommand{\bmu}		{\boldsymbol{\mu}}
\newcommand{\bnu}		{\boldsymbol{\nu}}
\newcommand{\bxi}		{\boldsymbol{\xi}}

\newcommand{\bphi}		{\boldsymbol{\phi}}

\global\long\def\mult#1#2{\mu_{#1}^{#2}}
\global\long\def\mul#1{\boldsymbol{\mu}_{#1}}
\global\long\def\xmul#1{\boldsymbol{x}_{#1}}
\global\long\def\xmult#1#2{x_{#1}^{#2}}
\global\long\def\Mmul#1{\boldsymbol{M}_{#1}}
\global\long\def\hmul#1{\boldsymbol{h}_{#1}}
\global\long\def\hmult#1#2{h_{#1}^{#2}}
\global\long\def\diff#1#2{\delta_{#1}^{#2}}
\global\long\def\dif#1{\boldsymbol{\delta}_{#1}}

\global\long\def\ymul#1{\boldsymbol{y}_{#1}}
\global\long\def\ymult#1#2{y_{#1}^{#2}}
\global\long\def\zmul#1{\boldsymbol{z}_{#1}}
\global\long\def\zmult#1#2{z_{#1}^{#2}}

\global\long\def\ev{\mathbb{E}}
\global\long\def\ab{\mathcal{A}}

\global\long\def\ga{\mathsf{A}}
\global\long\def\gc{\mathsf{C}}
\global\long\def\gg{\mathsf{G}}
\global\long\def\gt{\mathsf{T}}
\global\long\def\gx{\mathsf{X}}
\newcommand{\delt}[3]{D_{#2,#3}(#1)}

\newcommand{\abs}[1]{\left| #1 \right|}
\newcommand{\parenv}[1]{\left( #1 \right)}
\newcommand{\mathset}[1]{\left\{ #1 \right\}}
\newcommand{\ccap}{\mathsf{cap}}
\newcommand{\longversion}[1]{#1}
\newcommand{\explaination}[1]{}

\renewcommand{\le}{\leqslant}
\renewcommand{\leq}{\leqslant}
\renewcommand{\ge}{\geqslant}
\renewcommand{\geq}{\geqslant}

\begin{document}
\title{Evolution of $k$-mer Frequencies and Entropy in \\
 Duplication and Substitution Mutation Systems}
 
\author{Hao~Lou,~\IEEEmembership{Student~Member,~IEEE,} Farzad~Farnoud~(Hassanzadeh),~\IEEEmembership{Member,~IEEE,}\\ Moshe~Schwartz,~\IEEEmembership{Senior Member,~IEEE,} and~Jehoshua~Bruck,~\IEEEmembership{Fellow,~IEEE}%
\thanks{This paper was presented in part at ISIT 2018~\cite{lou2018} and ISIT 2015~\cite{farnoud2015b}.}%
\thanks{Hao Lou is with the Department
    of Electrical and Computer Engineering, 
    University of Virginia, Charlottesville, VA, 22903, USA, (email: hl2nu@virginia.edu).}%
\thanks{Farzad Farnoud (Hassanzadeh) is with the Department of Electrical and Computer Engineering, University of Virginia, Charlottesville, VA, 22903, USA, (email: farzad@virginia.edu).}%
  \thanks{Moshe Schwartz is with the Department
    of Electrical and Computer Engineering, Ben-Gurion University of the Negev,
    Beer Sheva 8410501, Israel
    (e-mail: schwartz@ee.bgu.ac.il).}%
  \thanks{Jehoshua Bruck is with the Department
    of Electrical Engineering, California Institute of Technology, Pasadena, CA 91125, USA (e-mail: bruck@paradise.caltech.edu).}%
}
\maketitle
\begin{abstract}
 Genomic evolution can be viewed as string-editing processes driven by mutations. An understanding of the statistical properties resulting from  these mutation processes is of value in a variety of tasks related to biological sequence data, e.g., estimation of model parameters and compression. At the same time, due to the complexity of these processes, designing tractable stochastic models and analyzing them are challenging. In this paper, we study two kinds of systems, each representing a set of mutations. In the first system, tandem duplications and substitution mutations are allowed and in the other, interspersed duplications. We provide stochastic models and, via stochastic approximation, study the evolution of substring frequencies for these two systems separately. Specifically, we show that $k$-mer frequencies converge almost surely and determine the limit set. Furthermore, we present a method for finding upper bounds on entropy for such systems.
\end{abstract}


\section{Introduction\label{sec: intro}}
\IEEEPARstart{D}{ue} to advances in DNA sequencing, vast amounts of biological sequence data are available nowadays. Developing efficient methods for the analysis and storage of this type of data will benefit from gaining a better mathematical understanding of the structure of these sequences. Biological sequences are formed by genomic mutations, which alter the sequence in each generation to create a new sequence in the next generation. These processes can be viewed as stochastic string editing operations that shape the statistical properties of sequence data.  


In this paper, our goal is to gain a better understanding of the evolution of sequences under random mutations. We represent the evolutionary process as a stochastic system in which an arbitrary initial string evolves through random mutation events. In such systems, we will study the evolution of the frequencies of words of length $k$, i.e., $k$-mers, as the sequence evolves. The analysis of $k$-mers has various applications, including identifying functions and evolutionary features~\cite{sievers2017}. Alignment-free sequence comparison also relies on $k$-mer frequencies~\cite{zielezinski2017}. Their analysis is also of interest because other statistical properties can be computed from $k$-mer frequencies. 

From an information-theoretic point of view, stochastic sequence generation process through mutation can be viewed as a \emph{source} of information. We study the entropy of such sources, which can be viewed as representing the complexity of sequences generated by the source. Sequence complexity measures, including entropy, have been used to determine the origin and/or the role of DNA sequences~\cite{orlov2004,farach1995,wan2003}, for example to classify protein-coding and non-coding regions of a genome. The entropy of a source also determines how well the sequences it produces can be compressed, an increasingly important problem given the growth of biological data. 

Several types of mutations exist, including substitution, duplication, insertion, and deletion. Substitution refers to changing a symbol in the sequence, e.g., $\mathsf{ACGTCT} \to \mathsf{ACG\underline{C}CT}$. Duplication mutations, where a segment of DNA (called the template) is
copied and inserted elsewhere in the genome, may be of the
\emph{tandem} or \emph{interspersed} type. In tandem duplication, the copy is inserted immediately after the template. For example, from $\mathsf{ACGTCT}$, we may obtain $\mathsf{AC\overline{GT}\underline{GT}CT}$, where the template is overlined and the copy is underlined. For interspersed duplication, there is generally no relationship between where the template is located and where the copy is inserted. As an example, two possibilities for $\ga\gg\gt\gt\gc$ after a single interspersed duplication are $\overline{\ga\gg\gt}\gt\underline{\ga\gg\gt}\gc$ and $\overline{\ga\gg}\underline{\ga\gg\gt}\overline{\gt}\gt\gc$. Our focus will be on duplication mutations, which are thought to play a critical role in generating new genetic material~\cite{ohno1970evolution}. 

Tandem duplication is generally thought to be caused by slipped-strand mispairings~\cite{mundy2004}, where during DNA synthesis, one strand in a DNA duplex becomes misaligned with the other. Tandem duplications and substitutions, along with other mutations, lead to tandem repeats, i.e., stretches of DNA in which the same pattern is repeated many times. Tandem repeats are known to cause important phenomena such as chromosome fragility~\cite{usdin2008a}. Interspersed duplications are caused by transposons, or ``jumping genes'', which are elements in the genome that can ``copy/paste'' themselves into different locations. Interspersed duplication is of interest as it leads to interspersed repeats, which form 45\% of the human genome~\cite{lander2001initial}.

We will analyze two systems involving the types of duplications discussed. The first system models a sequence evolving through tandem duplications and substitutions (TDS) and the second system represents interspersed duplications (ID). Along with duplications, other types of mutations occur. But for simplicity, our attention is limited to the aforementioned systems, and we leave more comprehensive analysis to future work. Furthermore, the significantly more complex effect of natural selection is not considered.

In TDS systems, in each step, i) a randomly chosen substring of the sequence is duplicated and inserted in tandem, or ii) a position is chosen at random and the symbol in that position is changed to one of the other symbols. In ID systems, a string evolves through random interspersed-duplication events, i.e., in each step, a random segment of the string is duplicated and then inserted in a random position in the string, independent of the position of the original segment. 
 
 
Our analysis starts by considering how $k$-mer frequencies evolve as mutations occur. To analyze their evolution, we use the stochastic-approximation method, which enables modeling of a discrete dynamic system by a corresponding continuous system described by  an ordinary differential equation (ODE). For the TDS model, our approach allows us to compute the limit for the frequency of any $k$-mer as a function of model parameters. We will then use these results to provide bounds on the entropy of sequences generated by tandem duplications and substitutions. For the ID model, we show that 
the frequencies of strings of length larger than one are, in the limit,
consistent with those of iid sequences; implying that in a certain
sense, a sequence evolving through interspersed duplication is unrecognizable
from an iid sequence. Note that an iid sequence has the maximum entropy
among sequences with a given symbol distribution. The structure of the limit set for $k$-mer frequencies in ID systems, however, leads to trivial upper bounds on the entropy. However, in certain cases these bounds are satisfied with equality.

In previous work, the related problem of finding the combinatorial capacity of duplication systems has been studied. The combinatorial capacity is related to entropy but is defined based on the size of the set of sequences that can be generated by the system, without considering their probabilities. The combinatorial capacity is studied by~\cite{farnoud2016,jain2017c}, for duplication systems (without allowing other types of mutations) and by~\cite{jain2017b} for systems with both tandem duplication and substitution. Compared to combinatorial capacity, entropy, which is studied in this paper, provides a more accurate measure of the complexity and compressibility of sequences generated by the system. For duplication systems and duplication/substitution systems, entropy has been studied by~\cite{elishco2016b,elishco2018}. While this work considers a wider range of systems, it only allows duplications involving single symbols. Furthermore, it does not study $k$-mer frequencies. The stochastic-approximation framework has been used for estimation of model parameters in tandem duplication systems~\cite{farnouda}. Estimating the entropy of DNA sequences has been studied in~\cite{schmitt1997,DNAEntropyDatacompression,farach1995}. However these works focus on estimating the entropy from a given sequence, rather than computing the entropy of a stochastic sequence generation system that models evolution. Duplication systems have also been studied in the context of designing error-correcting codes~\cite{jain2017duplication,dolecek2010repetition,chee2017deciding,lenz2017bounds}.

The rest of the paper is organized as follows. Notation and preliminaries
are given in the next section. In Section~\ref{sec: SA}, we present the framework for the application of stochastic approximation to our string-duplication systems. Section~\ref{sec: TDS}
contains the analysis of the evolution of $k$-mer frequencies in tandem duplication systems and the proof of entropy bounds. Section~\ref{sec: ID} is devoted to the analysis of $k$-mer frequencies in strings undergoing random interspersed duplications. We close the paper with concluding remarks in Section~\ref{sec: Con}.

\section{Notation and Preliminaries\label{sec: pre}}

For a positive integer $m$, let $[m]=\{1,\dotsc,m\}$. For a finite alphabet $\cA$, the set of all finite strings over $\cA$ is denoted $\cA^*$, and the set of all finite non-empty strings is denoted $\cA^+$. Also, let $\cA^k$ denote the set of $k$-mers, i.e., length-$k$ strings, over $\cA$. The elements in strings are indexed starting from 1, e.g., $\bs=s_1\dotsm s_{m}$, where $|\bs|=m$ is the length of $\bs$. For a string $\bu \in \cA^*$, $\bu_{i,j}$ denotes the length-$j$ substring of $\bu$ starting at $u_i$. Furthermore, the concatenation of two strings $\bu$ and $\bv$ is denoted by $\bu\bv$. For a non-negative integer $j$, and $\bu\in\cA^*$, $\bu^j$ is a concatenation of $j$ copies of $\bu$. Vectors and strings are denoted by boldface letters such as $\bx$, while scalars and symbols by normal letters such as $x$.

Consider an initial string $\bs_0$ and a process where in each step a random transform, or ``mutation'', is applied to $\bs_{n}$, resulting in $\bs_{n+1}$. To avoid the complications arising from boundaries, we assume the strings $\bs_n$ are circular, with a given origin and direction. Let the length of $\bs_n$ be denoted by $L_n$. To a duplication of length $\ell$, which may be tandem or interspersed depending on the model under study, we assign probability $q_{\ell}$. For TDS systems, in which substitutions are present, we denote the probability of substitution with $q_0$. For ID systems, we let $q_0=0$. The position of the template in duplication mutations is chosen at random among the $|\bs_n|$ possible options. For interspersed duplication, the position at which the copy is inserted is also chosen randomly. Furthermore, for substitution mutations, the position of the symbol that is substituted is chosen randomly. We assume there exists $M$ such that $q_{\ell}=0$ for all $\ell\geq M$. Hence, we have $\sum_{\ell=0}^{M-1}q_\ell=1$. 

For a string $\bu \in \cA^+$, denote the number of appearances of $\bu$ in $\bs_n$ as $\mu_{n}^{\bu}$, and its frequency as $x_{n}^\bu$, where $x_{n}^\bu = \mu_{n}^{\bu}/L_n$. For example, if $\bs_n = \ga\gc\gg\ga\gc$, then $\mu_{n}^{\ga\gc} = 2, x_{n}^{\ga\gc} = \frac{2}{5}$. Furthermore, for any set $U\subseteq \cA^+$, we define $\bmu_n = (\mu_n^{\bu})_{\bu\in U}$, and $\bx_n = (x_n^{\bu})_{\bu \in U}$. Thus $\bmu_n$ is a vector representing the number of appearances of $\bu\in U$ in the string $\bs$ at time $n$ and $\xmul n$ is the normalized version of $\mul n$. We let $\left\{ \mathcal{F}_{n}\right\} $
be the filtration generated by the random variables $\left\{ \xmul n,L_{n}\right\} $. 

Before proceeding to the analysis of $k$-mer frequencies, we present two results for the evolution of symbol frequencies (1-mers). These results can be viewed as extensions of results for P\'olya urn models. In such models, a random ball is chosen from an urn containing balls of different colors. The chosen ball is returned to the urn, along with a predetermined number of balls of the same color. It is known that the ratio of the balls of each color (equivalent to symbol frequencies) is a martingale and converges almost surely. While strings are more complex objects than urns, we describe similar results that are valid for any duplication process in which for each $i$, all $i$-substring of $\bs$ have the same chance of being duplicated. In particular, these results hold both for TDS systems with $q_0 = 0$ and for ID systems.
\begin{thm}
\label{thm:singleton1}In a duplication system with $q_0 = 0$, the random variables $\xmult na$, $a\in \cA$, are martingales and converge almost surely.\end{thm}
\begin{IEEEproof}
Suppose $a\in\ab$. We have 
\begin{align*}
\ev\left[\xmult{n+1}a|\mathcal{F}_{n}\right] & =\ev\left[\left.\frac{\mult{n+1}a}{L_{n+1}}\right|\mathcal{F}_{n}\right]=\ev\left[\left.\ev\left[\left.\frac{\mult{n+1}a}{L_{n+1}}\right|\mathcal{F}_{n},\ell\right]\right|\mathcal{F}_{n}\right]\\
 & =\ev\left[\left.\frac{\mult na+\ell\xmult na}{L_{n}+\ell}\right|\mathcal{F}_{n}\right]=\xmult na.
\end{align*}
We thus have $E\left[\xmult{n+1}a|\mathcal{F}_{n}\right]=\xmult na$
and so $\xmult na$ is a martingale. Since it is nonnegative, 
by the martingale convergence theorem, it converges almost
surely.
\end{IEEEproof}
The above theorem does not in fact require the distribution $q$ to
be constant and bounded. Under our assumption that $q$ is so, we can in addition obtain the following result
on the probability of $\xmult na$ deviating from its starting value.
\begin{thm}
\label{thm:singleton2}For all $a\in\mathcal{A}$ and $n\ge1$ we have
\[
\Pr\left(\left|\xmult na-\xmult 0a\right|\ge\lambda\right)\le2e^{-\lambda^{2}L_{0}^{2}/(2M^{4})}\ .
\]
\end{thm}
\begin{IEEEproof}
Since $q_{i}=0$ for $i\ge M$ and $i\le0$, $\frac{\mult{n-1}a}{L_{n-1}+M}\le\frac{\mult na}{L_{n}}\le\frac{\mult{n-1}a+M}{L_{n-1}+M}$.
Thus 
\[
-\frac{M\mult{n-1}a}{L_{n-1}(L_{n-1}+M)}\le\frac{\mult na}{L_{n}}-\frac{\mult{n-1}a}{L_{n-1}}\le\frac{M(L_{n-1}-\mult{n-1}a)}{L_{n-1}(L_{n-1}+M)}\ ,
\]
implying that 
\begin{align*}
\left|\xmult na-\xmult{n-1}a\right| & \le\frac{M\max\left\{ L_{n-1}-\mult{n-1}a,\mult{n-1}a\right\} }{L_{n-1}(L_{n-1}+M)}\\
 & \le\frac{M}{L_{n-1}+M}\le\frac{M}{L_{0}+n-1+M}\le\frac{K}{L_{0}+n}\ .
\end{align*}
Let $c_{n}=\frac{K}{L_{0}+n}$ so that $\left|\xmult na-\xmult{n-1}a\right|\le c_{n}$
and note that 
\begin{align*}
\sum_{i=1}^{n}c_{i}^{2} & =M^{2}\sum_{i=1}^{n}\frac{1}{\left(L_{0}+i\right)^{2}}\le M^{2}\int_{0}^{n}\frac{dt}{\left(L_{0}+t\right)^{2}}\\
 & =\frac{M^{2}}{L_{0}}-\frac{M^{2}}{L_{0}+n}=\frac{M^{2}n}{L_{0}\left(L_{0}+n\right)}\le\frac{M^{2}}{L_{0}}.
\end{align*}
By the Hoeffding-Azuma inequality, we have $\Pr\left(\left|\xmult na-\xmult 0a\right|\ge\lambda\right)\le2\exp\left(\frac{-\lambda^{2}L_{0}^{2}}{2M^{4}}\right)$.
\end{IEEEproof}
The preceding theorem implies that it is unlikely for the composition
of a long DNA sequence to change dramatically through random 
duplication events of bounded length. Such changes, if observed, are likely the result of context-dependent duplications or other biased mutations.

Unfortunately, this simple martingale argument does not extend to
$\xmult n\bu$ when $|\bu|>1$. Therefore, for analyzing such cases, we use the more
flexible technique of stochastic approximation as described in the
sequel. 


\section{Stochastic Approximation \\for Duplication Systems\label{sec: SA}}

In this section, we present an overview of the application of stochastic approximation to the analysis of duplication systems. 
By using stochastic approximation, our goal is to find out how $\xmul n$ changes with $n$ by finding
a differential equation whose solution approximates $\xmul n$.

We state a set of conditions that must be satisfied for our analysis.
Let $\ev_{\ell}\left[\ \cdot\ \right]$ denote the expected value
conditioned on the fact that the length of the duplicated substring
is $\ell$ and let $\dif{\ell}=\ev_{\ell}\left[\mul{n+1}|\mathcal{F}_{n}\right]-\mul n$.
In the case of substitution, we let $\ell=0$. We consider the following conditions. 

\begin{assumption}\label{asm:pk}There exists $M\in\mathbb{N}$ such
that $q_{i}=0$ for $i\ge M$.\end{assumption}

\begin{assumption}

\label{asm:bounded-delta}$\mul{n+1}-\mul n$, and thus $\dif{\ell}$,
are bounded.\end{assumption}

\begin{assumption}

\label{asm:bounded-x}$\xmul n$ is bounded.\end{assumption}

\begin{assumption}

\label{asm:delta-f-of-x}For each $\ell$, $\dif{\ell}$ is a function
of $\xmul n$ only, so we can write $\dif{\ell}=\dif{\ell}\left(\xmul n\right)$.\end{assumption}

\begin{assumption}

\label{asm:delta-lipshitz}The function $\dif{\ell}\left(\xmul n\right)$
is Lipschitz.

\end{assumption}

We assume~(A\ref{asm:pk}) holds. From this follows (A.\ref{asm:bounded-delta}) since for each $k$-mer, a mutation can create or eliminate a bounded number of occurrences. Additionally, (A.\ref{asm:bounded-x}) holds because each element of $\bx_n$ is between 0 and 1. The correctness of (A.\ref{asm:delta-f-of-x}) and (A.\ref{asm:delta-lipshitz}) will be shown for each system.

To understand how $\xmul n$ varies, our starting point is its difference
sequence $\xmul{n+1}-\xmul n$. We note that 
\begin{equation}
\xmul{n+1}-\xmul n=\ev\left[\xmul{n+1}-\xmul n|\mathcal{F}_{n}\right]+\left(\xmul{n+1}-\ev\left[\xmul{n+1}|\mathcal{F}_{n}\right]\right).\label{eq:1st-rel}
\end{equation}
For the first term of the right side of~(\ref{eq:1st-rel}), we have 
\begin{align}
\ev\left[\xmul{n+1}-\xmul n|\mathcal{F}_{n}\right] & =\sum_{\ell=0}^{M-1}q_{\ell}\left(\ev_{\ell}\left[\xmul{n+1}|\mathcal{F}_{n}\right]-\xmul n\right)\nonumber \\
 & =\sum_{\ell=0}^{M-1}q_{\ell}\left(\frac{\mul n+\dif{\ell}\left(\xmul n\right)}{L_{n}+\ell}-\frac{\mul n}{L_{n}}\right)\nonumber \\
 & =\frac{1}{L_{n}}\sum_{\ell=0}^{M-1}q_{\ell}\hmul{\ell}\left(\xmul n\right)\left(1+O\left(L_{n}^{-1}\right)\right)\nonumber \\
 & =\frac{1}{L_{n}}\hmul{}\left(\xmul n\right)\left(1+O\left(L_{n}^{-1}\right)\right),\label{eq:expected-change}
\end{align}
where $\bh_{\ell}(\bx_n) = \bdelta_\ell (\bx_n) - \ell \bx_n,$
$\bh(\bx_n) = \sum_{\ell = 0}^{M-1}q_{\ell}\bh_{\ell}(\bx_n)$,
and where we have used $1/\left(L_{n}+\ell\right)=\left(1+O\left(L_{n}^{-1}\right)\right)/L_{n}$,
which follows from the boundedness of $\ell$ (see (A\ref{asm:pk})).

Furthermore, for the second term of the right side of~(\ref{eq:1st-rel}),
we have
\begin{align}
\xmul{n+1}-\ev\left[\xmul{n+1}|\mathcal{F}_{n}\right] & =\frac{\mul{n+1}}{L_{n+1}}-\ev\left[\frac{\mul{n+1}}{L_{n+1}}|\mathcal{F}_{n}\right]\nonumber \\
 & =\frac{1+O\left(L_{n}^{-1}\right)}{L_{n}}\left(\mul{n+1}-\ev\left[\mul{n+1}|\mathcal{F}_{n}\right]\right)\nonumber \\
 & =\frac{1}{L_{n}}\left(1+O\left(L_{n}^{-1}\right)\right)M_{n+1}\label{eq:mart-diff},
\end{align}
where $\Mmul{n+1}=\mul{n+1}-\ev\left[\mul{n+1}|\mathcal{F}_{n}\right]$.
Note that $\Mmul n$ is a bounded martingale difference sequence.

From (\ref{eq:1st-rel}), (\ref{eq:expected-change}), and (\ref{eq:mart-diff}),
we find
\[
\xmul{n+1}=\xmul n+\frac{1}{L_{n}}\left(\hmul{}\left(\xmul n\right)+\Mmul{n+1}+O\left(L_{n}^{-1}\right)\right),
\]
where we have used the fact that $\hmul{}\left(\xmul n\right)\left(1+O\left(L_{n}^{-1}\right)\right)=\hmul{}\left(\xmul n\right)+O\left(L_{n}^{-1}\right)$.
This follows from the boundedness of $\hmul{}\left(\xmul n\right)$,
which in turn follows from the boundedness of $\dif{\ell}\left(\xmul n\right)$ for all $0\le\ell<M$. We note that $\bh$ determines the overall expected behavior of the system.

In the rest of the paper, the element of $\bdelta_{\ell}(\bx_{n})$ that corresponds to $\bu$ is denoted by $\delta^{\bu}_{\ell}(\bx_n)$. More precisely, $\delta_{\ell}^{\bu}(\bx_n) = \bbE_{\ell}[\mu_{n+1}^{\bu}-\mu_{n}^{\bu}|\cF_{n}]$. This notation also extends to $\bh$.

An additional condition requires $\sum_{n}1/\abs{\bs_n} = \infty$ and $\sum_{n}1/\abs{\bs_n}^2 < \infty$, which can be proven using the Borel-Cantelli lemma if $q_0 < 1$. Given these and our discussion above, the following theorem, which relates the discrete system describing $\xmul n$ to a continuous system, follows from the stochastic-approximation theorem~\cite[Theorem~2]{borkar2008stochastic}. 
\begin{thm}
\label{thm:borkar} The vector of $k$-mer frequencies 
$\xmul n$ converges almost surely to a compact connected
internally chain transitive invariant set of the ODE 
$
d\xmul t/dt=\hmul{}\left(\xmul t\right).\label{eq:ode}
$
\end{thm}
Note the dual use of the symbol $\xmul{}$ in the theorem; the meaning
is however clear from the subscript. We recall that~\cite[Theorem~2]{borkar2008stochastic} a set $A$ is an
\emph{invariant} set of an ODE $d\zmul t/dt=\boldsymbol{f}\left(\zmul t\right)$
if it is closed and $\zmul{t'}\in A$ for some $t'\in\mathbb{R}$
implies that $\zmul t\in A$ for all $t\in\mathbb{R}$. The invariant
set $A$ is \emph{internally chain transitive }with respect to the
ODE $d\zmul t/dt=\boldsymbol{f}\left(\zmul t\right)$, provided that
for every $\ymul{},\ymul{}'\in A$ and positive reals $T$ and $\epsilon$,
there exist $N\ge1$ and a sequence $\ymul 0,\dotsc,\ymul N$ with
$\ymul i\in A$, $\ymul 0=\ymul{}$, and $\ymul N=\ymul{}'$ such
that for $0\le i<n$, if $\zmul 0=\ymul i$, then for some $t\ge T$,
$\zmul t$ is in the $\epsilon$-neighborhood of $\ymul{i+1}$.

\section{Tandem Duplication with Substitution \label{sec: TDS}}
In this section, we study the behavior of a system that allows tandem duplication and substitution mutations. First, we will determine the limits of the frequencies of $k$-mers. Then, after presenting a theorem relating the limits to entropy, we find bounds on the entropy of these systems. 

Let $U = \cA^k$, so $\bmu_n$ is the vector of all $k$-mer occurrences, and $\bx_n$ is the vector of all $k$-mer frequencies. From Section~\ref{sec: SA} we know that we can use the differential equation $d\bx_t/dt = \bh(\bx_t)$ to determine the limit of $k$-mer frequencies. To find the differential equation, in Theorem~\ref{thm:TDS_SA}, we determine $\delta_{\ell}^{\bu}(\bx_n)$ for $\ell$ with $q_{\ell}>0$ and $\bu\in U$, where it can be observed that  (A.\ref{asm:delta-f-of-x}) and (A.\ref{asm:delta-lipshitz}) hold in our model


In the next subsection, we will give some necessary definitions. We will then prove that $\delta_{\ell}^{\bu}(\bx_n)$ is a linear function of $\bx_n$, which leads to a linear first-order differential equation. This linear form facilitates determining the asymptotic behavior of the $k$-mer frequencies. We will then show that the entropy of stochastic string systems can be related to the capacity of semiconstrained systems defined by the limit set of the $k$-mer frequencies. Leveraging the simple form of the limits for systems with tandem duplications and substitutions, we will provide bounds on the entropy of these systems. 

\subsection{Definitions\label{subsec: def_tds}}
The following definitions will be useful for finding $\bdelta_{\ell}(\bx_n)$. For $\bu\in\cA^*$ and $m\in\bbN^{+}$, define $\bphi_{m}(\bu)$ to be a sequence of length $|\bu|$ whose $i$th element is determined by whether the symbol in position $i$ of $\bu$ equals the symbol in position $i-m$. More specifically, the $i$th element of $\bphi_{m}(\bu)$ is
\begin{align*}
    \bphi_{m}(\bu)_i = 
	\begin{cases}
	0, \qquad &m+1\leq i \leq \abs{\bu}, u_i = u_{i-m}\\
	\gx, \qquad &\text{otherwise}\\ 
	\end{cases}
\end{align*}
where $\gx$ is a dummy variable. Only the positions of `0's in $\bphi_{m}(\bu)$ are of importance to us. Let the lengths of the maximal runs of `0's immediately after the initial $\gx^m$ and at the end of $\bphi_{m}(\bu)$ be denoted by $ l_m^{\bu} $ and $r_m^\bu$, respectively. Note that either of $l_m^{\bu}$ or $r_m^{\bu}$ may be equal to 0. If $\bphi_{m}(\bu)=\gx^m0^{\abs{\bu}-m}$, then $ l_m^{\bu}=r_m^{\bu} = \abs{\bu}-m $. Otherwise, we have $\bphi_{m}(\bu)= \gx^m0^{l_m^{\bu}}Y0^{r_m^{\bu}}$, for some $Y$ that starts and ends with $\gx$. For example, for $\cA = \{\ga,\gc,\gg,\gt\}$, we may have 
\begin{alignat*}{16}
    \bu&=           & &\ga & &\gc & &\ga & &\ga & &\gc & &\gc & &\ga & &\gc & &\gc & &\ga & &\ga & &\gc & &\ga & &\ga & &\gc,\\
    \bphi_3(\bu)&=   & &\gx & &\gx & &\gx & \,&0   & \,&0\,   & &\gx & \,&0   & &\,0   & &\,0   & \,&0\,   & &\gx & \,&0   & &\,0   & &\,0   & &\,0, 
\end{alignat*}
and $l_m^\bu=2,$ and $r_m^\bu=4$. Note that a duplication of length $m$ is equivalent to inserting $m$ 0s into $\bphi_m(\bu)$. For example, $\bu$ may come from $\bu' = \ga\gc\ga\overline{\ga\gc\gc}\ga\ga\gc\ga\ga\gc$ after a length 3 tandem duplication with the overlined substring as a template and $\bphi_3(\bu)$ can be viewed as the result of inserting three 0s into $\bphi_3(\bu') = \gx\gx\gx 0 0 \gx \overline{0\gx} 0000$ between the overlined two symbols.

To enable us to succinctly represent the results, we define several functions. These functions relate $\bu$ to the frequencies of other substrings that can generate $\bu$ via appropriate duplication events. 

Consider the sequence $\bu=\ga\gc\ga\gc\ga\gg\ga\gg$, for which $\bphi_2(\bu)=\gx\gx000\gx00$. This string can be created through duplications of length 2 from $\ga\gc\ga\gg\ga\gg$ (in two ways) and from $\ga\gc\ga\gc\ga\gg$. These correspond to runs of $0$ of length $2$ in $\bphi_2(\bu)$. For a string $\bu$ and positive integers $m$ and $z\in|\bu|-m+1$, let $\delt{\bu}{z}{m}$ be the sequence $\bu_{1,z-1}\bu_{z+m,\abs{\bu}-z-m+1}$ obtained from $\bu$ by removing symbols in positions $z,\dotsc,z+m-1$. For example, $\delt{\ga\gc\ga\gc\ga\gg\ga\gg}{4}{2}=\ga\gc\ga\gg\ga\gg$. Define 
    \begin{align*}
        G_{m}^{\bu}(\bx) = \sum_{z}x^{\delt{\bu}{z}{m}},
    \end{align*}
where the sum is over all $z$ that are the indices of the start of (not necessarily maximal) runs of $0$s in $\bphi_m(\bu)$, i.e., $\left(\bphi_{m}(\bu)\right)_{z,m}=0^{m}$. For $\bu = \ga\gc\ga\ga\gc\gc\ga\gc\gc\ga$, $G_3^{\bu}(\bx) = 2x^{\ga\gc\ga\ga\gc\gc\ga}$. There is a slight abuse of notation in the definition of $G$ above (as well as the definitions of $F$ and $M$ below). While the argument of $G$ is $\bx=\left(x^\bv\right)_{\bv\in\mathcal{A}^k}$, on the right side, $x^\bw$ for sequences $\bw$ with $|\bw|<k$ may appear. We note however that $x^\bw$ can be obtained from $\bx$ by summing over the elements of $\bx$ corresponding to strings that include $\bw$ as a prefix.

New occurrences of $\bu$ can also be generated from strings that are not of the form $\delt{\bu}{z}{m}$. For example, consider the sequence $\bu = \ga\gc\gg\ga\gc\gt\gg$, for which $\bphi_3(\bu) = \gx\gx\gx 00 \gx\gx $. This sequence can be created through a length-3 tandem duplication from $\overline{\gc\gg\ga}\gc\gt\gg$ and $\overline{\gg\ga\gc}\gt\gg$, where the part that is to be duplicated is overlined. The following definitions will be of use in the analysis of this type of duplication. 
\begin{align*}
    F^{\bu}_{m,l}(\bx)&=\sum\limits_{i = 1}^{\min(l_m^{\bu},m-1)}\!\!\!\!x^{\bu_{i+1,\abs{\bu}-i}},\\
    F^{\bu}_{m,r}(\bx)&=\sum\limits_{i=1}^{\min(r_m^{\bu},m-1)}\!\!\!\!x^{\bu_{1,\abs{\bu}-i}}. 
\end{align*}
In the special case where $\bphi_m(\bu)=\gx^m0^{|\bu|-m}$ and $\abs{\bu} \leq 2m-2$, we will benefit from defining
\begin{align*}
    M_{m}^{\bu}(\bx) =
    \sum\limits_{b = \abs{\bu}-m+1}^{m-1}x^{\bu_{b+1,m-b}\bu_{1,b}}.
\end{align*}
We let $M_{m}^{\bu}(\bx) =0$ if $\bphi_m(\bu)\neq\gx^m0^{|\bu|-m}$.


Lastly, we use $\cB_1(\bu)$ to denote set of strings at Hamming distance 1 from $\bu$. For example, for $\bu=\ga\gc$, $\cB_{1}(\bu) = \{\gg\gc, \gc\gc, \gt\gc, \ga\ga, \ga\gg, \ga\gt\}$. Also for $\bu,\bv\in\cA^*$, the indicator function $I(\bu,\bv)$ equals 1 if $\bu=\bv$ and equals 0 otherwise.

\subsection{Evolution of $k$-mer Frequencies\label{subsec: main_tds}}\label{sec:tandem-sub}
We first find $\bdelta_{\ell}(\bx)=\left(\delta_{\ell}^{\bu}(\bx)\right)_{\bu\in U}$ for $\ell>0$ (duplication) and then for $\ell=0$ (substitution). 

When analyzing $\delta_{\ell}^{\bu}(\bx)$, we only consider substrings $\bu$ of length $\abs{\bu}>\ell$, which simplifies the derivation. The frequencies of shorter substrings can be found by summing over the frequencies of longer substrings. 

\begin{thm}\label{thm:TDS_SA} For an integer $\ell>0$ and a string $\bu = u_1u_2\dotsm u_{k} $, if $ \ell+1\leq k<2\ell $, then
\begin{align*}
    \delta_{\ell}^{\bu}(\bx)  = F_{\ell,l}^{\bu}(\bx) + F_{\ell,r}^{\bu}(\bx) + M_{\ell}^{\bu}(\bx)- (k-1-\ell)x^{\bu},
\end{align*}
and if $k\ge2\ell$,
\begin{align*}
    \delta_{\ell}^{\bu}(\bx)  = F_{\ell,l}^{\bu}(\bx) + F_{\ell,r}^{\bu}(\bx) + G_{\ell}^{\bu}(\bx) - (k-1-\ell)x^{\bu}.
\end{align*}
\end{thm}
Before proving the theorem, we present two examples for $\ell=3$ and $\cA = \{\ga,\gc,\gg,\gt\}$:
\begin{align*}
    \delta_3^{\ga\gc\gg\ga}(\bx) &= x^{\ga\gc\gg} + x^{\gc\gg\ga} +x^{\gg\ga\gc} \\
    \delta_3^{\ga\gc\gg\ga\gc\gg\ga\gc}(\bx) &= 3x^{\ga\gc\gg\ga\gc}+x^{\ga\gc\gg\ga\gc\gg} +x^{\ga\gc\gg\ga\gc\gg\ga}+\\ 
    &\qquad x^{\gg\ga\gc\gg\ga\gc}+x^{\gc\gg\ga\gc\gg\ga\gc}-4x^{\ga\gc\gg\ga\gc\gg\ga\gc}.
\end{align*}
\begin{IEEEproof}
Suppose  a duplication of length $ \ell $ occurs in $ \bs_n $, resulting in $ \bs_{n+1} $. The number of occurrences of $\bu$ may change due to the duplication event. To study this change, we consider the $k$-substrings of $\bs_n$ that are eliminated (do not exist in $\bs_{n+1}$) and the $k$-substrings of $\bs_{n+1}$ that are new (do not exist in $\bs_n$). Any new $k$-substring must intersect with both the template and the copy in $\bs_{n+1}$. Likewise, an eliminated $k$-substring must include symbols on both sides of the template in $\bs_n$, i.e., the template must be a strict substring of the $k$-substring that includes neither its leftmost symbol nor its rightmost symbol.

As an example, suppose 
\begin{align}
    \bs_n&=\bv\ga\gc\gg\gt\ga\gg\ga\gt\bw, &
    \bs_{n+1}&=\bv\ga\gc\gg\overline{\gt\ga\gg}\underline{\gt\ga\gg}\ga\gt\bw, \label{ex:sn}
\end{align}
where $\ell=3$, the (new) copy is underlined and the template is overlined, and $ \bv,\bw \in \cA^* $. Let $k=5$, the new $5$-substrings are $\gg\gt\ga\gg\gt$, $\gt\ga\gg\gt\ga$, $\ga\gg\gt\ga\gg$, $\gg\gt\ga\gg\ga$
and the eliminated substring is $\gg\gt\ga\gg\ga$. Note that here the two $\gg\gt\ga\gg\ga$ substrings are counted as different. Formally, let
\begin{align*}
    \bs_n & = a_1\dotsm a_i a_{i+1}\dotsm a_{i+\ell}a_{i+\ell+1}\dotsm a_{|s_n|},\\
    \bs_{n+1} &= a_1\dotsm a_ia_{i+1}\ldots a_{i+\ell}a_{i+1}\ldots a_{i+\ell}a_{i+\ell+1}\ldots a_{|s_n|},
\end{align*}
where the substring $a_{i+1}\dotsm a_{i+\ell}$ is duplicated. The new $k$-substrings created in $ \bs_{n+1} $ are
\begin{align*}
	\by_b = a_{i+\ell+1-b}a_{i+\ell+2-b}\ldots a_{i+\ell}a_{i+1}a_{i+2}\ldots a_{i+k-b},
\end{align*}
for $1\leq b \leq k-1$. Note that we have defined $\by_b$ such that the first element of the copy, $a_{i+1}$, is at position $b+1$ in $\by_b$. The $k$-substrings eliminated from $\bs_n$ are $    a_{i-c+1}\dotsm a_{i+k-c}$, for $1\le c\le k-\ell-1$.


\begin{figure}
	\begin{center}
		\begin{tikzpicture}[
		node distance = 0cm,
		gn/.style={minimum width=.3cm, minimum height=0.5cm, shape=rectangle,draw,font=\small,inner sep=0mm},
		tmpl/.style={pattern=north east lines, pattern color=green},
		copy/.style={pattern=north west lines, pattern color=red},
		mu/.style={},
		]
		\node (lbl) [gn, draw=none] {$\bs_{n+1}$};
		\node (pre) [right = of lbl, xshift=.3cm, gn,minimum width=1.5cm] {};
		\node (tmpl) [right = of pre, gn,tmpl,minimum width=1cm] {};
		\node (copy) [right = of tmpl, gn, minimum width = 1cm, copy]{};
		\node (post) [right = of copy, gn, minimum width = 1.5cm]{};
		
		\node (case1) [below = of lbl, yshift = -.1cm, gn, draw=none] {Case 1};
		\node (mu1) [right = of case1, xshift = 2cm, gn, minimum width = 2.8cm] {$\bu$};		
		\node (case2) [below = of case1, yshift = -.1cm, gn, draw=none] {Case 2};
		\node (mu2) [right = of case2, xshift = 1.4cm, gn, minimum width = 2.8cm] {$\bu$};
		
		\node (case3) [below = of case2, yshift = -.1cm, gn, draw=none] {Case 3};
		\node (mu3) [right = of case3, xshift = .8cm, gn, minimum width = 2.8cm] {$\bu$};

		\node (case4) [above = of lbl, yshift = .1cm, gn, draw=none] {Case 1};
		\node (mu4) [right = of case4, xshift = 2.4cm, gn, minimum width = 1.5cm] {$\bu$};		
		\node (case5) [above = of case4, yshift = .1cm, gn, draw=none] {Case 2};
		\node (mu5) [right = of case5, xshift = 1.9cm, gn, minimum width = 1.5cm] {$\bu$};
		
		\node (case6) [above = of case5, yshift = .1cm, gn, draw=none] {Case 3};
		\node (mu6) [right = of case6, xshift = 1.4cm, gn, minimum width = 1.5cm] {$\bu$};
		
		\node (dummy) [left = of pre, gn, draw=none, xshift=1mm, fill=white]{};
		\node (dummy) [right = of post, gn, draw=none, xshift=-1mm, fill=white]{};
		\end{tikzpicture}
	\end{center}
	\caption{Possible cases for new occurrences of $\bu$ in $\bs_{n+1}$. Cases above and below $\bs_{n+1}$ correspond to $\ell+1\le k<2\ell$ and $k\ge2\ell$, respectively. The hatched boxes, from left to right, are the template and the copy.}\label{fig:musn1}
\end{figure}
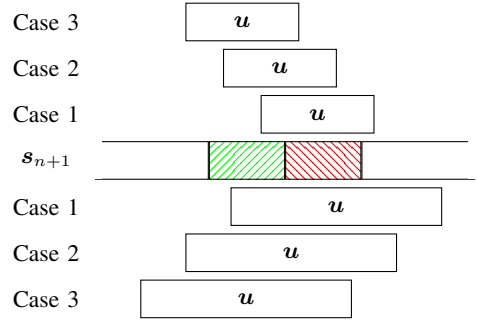

For a given $\bu$, let $Y_b$ denote the indicator random variable for the event that $\by_b=\bu$, that is, the duplication creates a new occurrence of $ \bu $ in $ \bs_{n+1}$ in which the first symbol of the copy is in position $b+1$. In example denoted by~(\ref{ex:sn}), if $\bu=\gt\ga\gg\gt\ga$, then $\by_3=\bu$ and thus $Y_3=1$.


Furthermore, let $W$ denote the number of occurrences of $\bu$ that are eliminated. We have
\begin{align*}
    \delta_{\ell}^{\bu}(\bx) &= \Big(\sum\limits_{b=1}^{k-1}\bbE_{\ell}[Y_b|\mathcal{F}_n]\Big) - \bbE_{\ell}[W|\mathcal{F}_n] \\
    &= \Big(\sum\limits_{b=1}^{k-1}\bbE_{\ell}[Y_b|\mathcal{F}_n]\Big) - (k-\ell-1)x^{\bu},
\end{align*}
where the second equality follows from the fact that each of the $k-\ell-1$ eliminated $k$-substrings are equal to $\bu$ with probability $x^{\bu}$.

To find $ \delta_k^{\bu} $, it suffices to find $ \bbE_{\ell}[Y_b|\mathcal{F}_n]$, or equivalently, $\Pr(Y_b=1|\mathcal{F}_n,\ell) $. We consider different cases based on the value of $ b $, which determines how $\bu$ overlaps with the template and the copy. These cases are illustrated in Figure~\ref{fig:musn1} and are considered in Lemmas~\ref{lem:case1}--\ref{lem:case3}. 

From Lemma~\ref{lem:case1}, we have
\begin{align}
    \sum_{b=1}^{{\min(\ell-1,k-\ell)}}\!\!\!\!\!\!\!\!\bbE_{\ell}[Y_b|\mathcal{F}_n]
    &=\sum_{b=1}^{\min(\ell-1,k-\ell)} \!\!\!\!\!x^{\bu_{b+1,k-b}}\eql{\bu_{1,b}}{\bu_{1+\ell,b}}\nonumber\\
    &=\sum_{b=1}^{\min(\ell-1,k-\ell)}\!\!\!\!\! x^{\bu_{b+1,k-b}}\eql{\bphi_{\ell}(\bu)_{\ell+1,b}}{0^b}\nonumber\\
    &=\sum_{b=1}^{\min(\ell-1,k-\ell,l_\ell^{\bu})} \!\!\!\!\!x^{\bu_{b+1,k-b}}\nonumber\\
    &=\sum_{b=1}^{\min(\ell-1,l_\ell^{\bu})} \!\!\!x^{\bu_{b+1,k-b}}\nonumber\\
    &=F^{\bu}_{\ell,l}(\bx), \label{eq:lm5}
\end{align}
where the fourth equality follows from the fact that $l_\ell^{\bu}\le k-\ell$.

Similarly, using Lemma~\ref{lem:case3}, it can be shown that
\begin{align}
    &\sum_{b=\max{(k-\ell+1,\ell)}}^{k-1}\bbE_{\ell}[Y_b|\mathcal{F}_n]=F_{\ell,r}^{\bu}(\bx), \label{eq:lm7}
\end{align}
\explaination{\begin{align*}
    &\sum_{b=\max{(k-\ell+1,\ell)}}^{k-1}\bbE_{\ell}[Y_b|\mathcal{F}_n]\\
    &=\sum_{b = \max{(k-\ell+1,\ell)}}^{k-1}x^{\bu_{1,b}}\eql{\bu_{b-\ell+1,k-b}}{\bu_{b+1,k-b}}\\
    &=\sum_{b = \max{(k-\ell+1,\ell)}}^{k-1}x^{\bu_{1,b}}\eql{\bphi_{\ell}(\bu)_{b+1,k-b}}{0^{k-b}}\\
    &=\sum_{b = \max{(k-\ell+1,\ell,k-r_\ell^{\bu})}}^{k-1}x^{\bu_{1,b}}\\
    &=\sum_{b = \max{(k-\ell+1,k-r_\ell^{\bu})}}^{k-1}x^{\bu_{1,b}}\\
    &=\sum_{i =1}^{\min{(r_\ell^{\bu},\ell-1)}}x^{\bu_{1,k-i}}\\
    &=F_{\ell,r}^{\bu}(\bx),
\end{align*}
where the fourth equality follows from $r_\ell^\bu\le k-\ell$ and the fifth equality comes from setting $i = k-b$.}

To complete the proof, we need to show that $\ev_{\ell}[Y_b|\cF_n]$ summed over the range $\min(\ell,k-\ell+1)\le b\le \max(k-\ell,\ell-1)$ reduces to $G_{\ell}^{\bu}(\bx)$ or $M_{\ell}^{\bu}(\bx)$ as appropriate.

From Lemma~\ref{lem:case2}, if $ \ell+1\leq k\le2\ell-2$, then
\begin{align}
    &\sum_{b=\min{(\ell,k-\ell+1)}}^{\max{(k-\ell,\ell-1)}} \bbE_{\ell}[Y_b|\mathcal{F}_n]
    =\sum_{b=k-\ell+1}^{\ell-1} \bbE_{\ell}[Y_b|\mathcal{F}_n]\nonumber\\
    &=\sum_{b = k-\ell+1}^{\ell-1}x^{\bu_{b+1,\ell-b}\bu_{1,b}}\eql{\bu_{1,k-\ell}}{\bu_{\ell+1,k-\ell}}\nonumber\\
    &=\sum_{b = k-\ell+1}^{\ell-1}x^{\bu_{b+1,\ell-b}\bu_{1,b}}\eql{\bphi_{\ell}(\bu)_{\ell+1,k-\ell}}{0^{k-\ell}} \nonumber\\
    &=
    M_{\ell}^{\bu}(\bx), \label{eq:lm6.1}
\end{align}

and if $k = 2\ell-1$, also
\begin{align}
    \sum_{b=\min{(\ell,k-\ell+1)}}^{\max{(k-\ell,\ell-1)}} \bbE_{\ell}[Y_b|\mathcal{F}_n] = 0 = M_{\ell}^{\bu}(\bx).\label{eq:lm6.2}
\end{align}

Finally, if $k\ge 2\ell$, from the same lemma, we find
\begin{align}
    &\sum_{b=\min{(\ell,k-\ell+1)}}^{\max{(k-\ell,\ell-1)}} \bbE_{\ell}[Y_b|\mathcal{F}_n]
    =\sum_{b=\ell}^{k-\ell} \bbE_{\ell}[Y_b|\mathcal{F}]\nonumber\\
    &=\sum_{b = \ell}^{k-\ell}x^{\bu_{1,b-\ell}\bu_{b+1,k-b}}\eql{\bu_{b-\ell+1,\ell}}{\bu_{b+1,\ell}}\nonumber\\
    & = \sum_{b = \ell}^{k-\ell}x^{\bu_{1,b-\ell}\bu_{b+1,k-b}}\eql{\bphi_{\ell}(\bu)_{b+1,\ell}}{0^{\ell}}
    =G_k(\bu),\label{eq:lm6.3}
\end{align}
where the last step follows from the definition of $G_k$. 

Summing over the expressions provided by~(\ref{eq:lm5})-(\ref{eq:lm6.3}) provides the desired result.
\end{IEEEproof}

\begin{lem}[Case 1]\label{lem:case1}
For $1\le b < \min(\ell,k-\ell+1)$,
\[\bbE_{\ell}[Y_b|\cF_n]=x^{\bu_{b+1,k-b}}\eql{\bu_{1,b}}{\bu_{1+\ell,b}}.\]
\end{lem}
\begin{IEEEproof}
For $1\leq b < \min(\ell,k-\ell+1)$ (regardless of whether $k\geq 2\ell$ or $k < 2\ell$), the new occurrences of $ \bu $ always contain some (but not all) of the template and all of the new copy. This scenario is labeled as Case 1 in Figure~\ref{fig:musn1}.
 
Suppose $Y_b=1$. Since the copy and the template are identical, elements of $\bu$ that coincide with the same positions in these two substrings must also be identical. So a necessary condition for $Y_b =1$ is
\begin{align*}
    \bu_{1,b} = \bu_{1+\ell,b}.
\end{align*}
Assume this condition is satisfied. Then $Y_b=1$ if and only if the sequence starting at the beginning of the template in $\bs_n$ is equal to $\bu_{b+1,k-b}$, which has probability $x^{\bu_{b+1,k-b}}$. 
As an example for $k\ge 2\ell$, consider
\begin{align*}
	 \bs_n &= \bv1234567\bw, \\
	 \bs_{n+1} &= \bv12\overline{3\underline{123}456}7\bw , \text{ where }  Y_b =1 \text{ for }  b=1, \\
	 u_1 &= u_4 = 3,
\end{align*}
where $\bv,\bw\in\cA^*$, $\bu$ is overlined, and the copy is underlined. Note that $\bs_n$ contains $\bu_{b+1,k-b}=123456$. For $k<2\ell$, consider
\begin{align*}
	 \bs_n &= \bv1234\bw, \\
	 \bs_{n+1} &= \bv12\overline{3\underline{123}4}\bw, \text{ where }  Y_b =1  \text{ for }  b =  1,\\
	u_1 &= u_4  = 3.
\end{align*}
\end{IEEEproof}
\begin{lem}[Case 2]\label{lem:case2}
Suppose $\min(\ell,k-\ell+1)\le b < \max(k-\ell+1,\ell)$. If $k\ge2\ell$, then
\[\bbE_{\ell}[Y_b|\cF_n] = x^{\bu_{1,b-\ell}\bu_{b+1,k-b}}\eql{\bu_{b-\ell+1,\ell}}{\bu_{b+1,\ell}},\]
and if $\ell+1\le k\le2\ell-2$, then
\[\bbE_{\ell}[Y_b|\cF_n] = x^{\bu_{b+1,\ell-b}\bu_{1,b}}\eql{\bu_{1,k-\ell}}{\bu_{\ell+1,k-\ell}}.\]
\end{lem}
\longversion{In this Case 2, $\bu$ either i) contains both the template and the copy completely, or ii) intersects with both but contains neither. }Note that this case cannot occur if $k=2\ell-1$.
\begin{IEEEproof}
First, assume $k\ge 2\ell$. The condition on $b$ translates to $\ell\le b<k-\ell+1$ and the new occurrence of $ \bu $ contains both the template and the copy. This is labeled as Case 2 in Figure~\ref{fig:musn1} (below $\bs_{n+1}$). With the same logic as in Case 1, it is clear that we need
\begin{center}
	$\bu_{b-\ell+1,\ell} = \bu_{b+1,\ell},$
\end{center}

Assuming this condition is satisfied, we have $ Y_b =1$ if and only if the substring $ \bu_{1,b-\ell}\bu_{b+1,k-b}$ occurs in $ \bs_n $ at a certain position, which occurs with probability $x^{\bu_{1,b-\ell}\bu_{b+1,k-b}}$. 

For example, consider
\begin{align*}
	 \bs_n &= \bv412356\bw, \\
	 \bs_{n+1} &= \bv\overline{4123\underline{123}5}6\bw, \text{ where }  Y_b = 1  \text{ for }  b=4, \\
	 \bu_{2,3} &= \bu_{5,3} =123. 
\end{align*}


Now suppose $\ell+1\le k\le 2\ell-2$. The condition on $b$ from the statement of the lemma is $ k-\ell+1 \le b < \ell$. The new occurrence of $ \bu $ contains some (but not all) of the elements of the template and some (but not all) of the elements of the copy, as illustrated in Figure~\ref{fig:musn1}, Case 2, above $\bs_{n+1}$. The following constraint on $ \bu $ must hold
\begin{center}
	$ \bu_{1,k-\ell} = \bu_{\ell+1,k-\ell}, $
\end{center} 
implying that $\phi_{\ell}(\bu) = \gx^\ell 0^{k-\ell} $. For example, consider
\begin{align*}
	 \bs_n &=\bv123\bw,\\
	 \bs_{n+1} &= \bv1\overline{23\underline{12}}\underline{3}\bw, \text{ where } Y_b = 1 \text{ for }  b=2, \\
   u_1 &= u_4 = 2.  
\end{align*}
We have $ Y_b =1 $ iff the sequence starting at the beginning of the template in $ \bs_n $ is equal to $ \bu_{b+1,\ell-b}\bu_{1,b} $, which has probability $x^{\bu_{b+1,\ell-b}\bu_{1,b}}$.

\end{IEEEproof}
\begin{lem}[Case 3]\label{lem:case3}
For $\max{(k-\ell+1,\ell)}\le b\le k-1$,
\[\bbE_{\ell}[Y_b|\cF_n]=x^{\bu_{1,b}}\eql{\bu_{b-\ell+1,k-b}}{\bu_{b+1,k-b}}.\]
\end{lem}
\begin{IEEEproof}
For $\max{(k-\ell+1,\ell)}\le b\le k-1$ (regardless of whether $k\geq 2\ell$ or $k < 2\ell$), the new occurrence of $ \bu $ contains the template and some (but not all) of the elements of the copy. This is labeled as Case 3 in Figure~\ref{fig:musn1}. The constraint on $ \bu $ is 
\begin{center}
	$ \bu_{b-\ell+1,k-b} = \bu_{b+1,k-b}. $
\end{center}
As examples, consider
\begin{align*}
	\bs_n&=\bv456123\bw,\\
	\bs_{n+1} &= \bv\overline{456123\underline{1}}\underline{23}\bw , \text{ where }  Y_b =1  \text{ for }  b=6, \\
	u_4 &= u_7 =1, 
\end{align*}
for $k \geq 2\ell$, and 
\begin{align*}
	 \bs_n &= \bv4123\bw, \\
	 \bs_{n+1} &= \bv\overline{4123\underline{1}}\underline{23}\bw, \text{ where } Y_b = 1  \text{ for }  b = 4, \\
	u_2 &= u_5 = 1,
\end{align*}
for $ \ell<k <2\ell$.

We have $ Y_b=1 $ if and only if $ \bu_{1,b} $ occurs in $\bs_n $ at a certain position, which has probability $x^{\bu_{1,b}}$.
\end{IEEEproof}
\begin{thm}\label{thm:k0}
For a string $\bu$ of length $k$, we have
\begin{align*}
    \delta_0^\bu(\bx) = \frac{1}{|\cA|-1}\sum_{\bv\in\cB_1(\bu)}x^\bv-kx^\bu.
\end{align*}
\end{thm}
Before proving the theorem, we give an example for $\cA = \{1,2,3\}$:
\[
    \delta_0^{123}(\bx) = \frac{1}{2}(x^{223} + x^{323}+x^{113} + x^{133}
    +x^{121} + x^{122}) - 3x^{123}
\]

\begin{IEEEproof}
A new occurrence of $\bu$ results from an appropriate substitution in some $\bv\in\cB_1(\bu)$, which has probability $x^{\bv}/(|\cA|-1)$. On the other hand, an occurrence of $\bu$ is eliminated if a substitution occurs in any of its $k$ positions. So the expected number occurrences that vanish is $kx^{\bu}$.
\end{IEEEproof}

\subsection{ODE and the Limits of Substring Frequencies\label{ODE_tds}}
Theorems~\ref{thm:TDS_SA} and~\ref{thm:k0} provide expressions for $\bdelta_{\ell}(\bx)$ for $0\le \ell\le M-1$. With these results in hand, we can formulate an ordinary differential equation (ODE) whose limits are the same as those of the substring frequencies of interest, $\bx = \left(x^\bu\right)_{\bu\in \cA^k}$, where $k\ge M$. 

We first show that $\delta_\ell ^{\bu}(\bx)$ can be written as a linear combination of the elements of $\bx$, i.e., a linear combination of $x^{\bv},\bv\in\cA^k$. To see this, note that on the right side in expressions for $\delta_{\ell}^\bu$ in Theorems~\ref{thm:TDS_SA} and~\ref{thm:k0}, terms of the form $x^{\bw}$ appear where $|\bw|\le k$. We can replace $x^{\bw}$ with $\sum_{\bv}x^{\bv}$, where the summation is over all strings $\bv$ of length $k$ such that $\bw$ is a prefix of $\bv$. 
For example, consider the alphabet $\{1,2,3\}$ and $k=3$. From Theorem~\ref{thm:TDS_SA}, we have \begin{align*}
     \delta_2^{121}(\bx)&=x^{12}+x^{21} \\
     &= x^{121} + x^{122}+ x^{123}+ x^{211}+ x^{212}+ x^{213}.
 \end{align*}

For $0\le \ell <M$, let $A_\ell$ be the matrix satisfying $\bdelta_\ell(\bx)-\ell \bx=A_\ell \bx$. Based on the argument above, such a matrix exists and can be computed from Theorems~\ref{thm:TDS_SA} and~\ref{thm:k0}. Furthermore, let 
\begin{equation}\label{eq:Adef}
    A=\sum_{\ell=0}^{M-1} q_{\ell} A_{\ell}.
\end{equation} 
Note that $\bh_{\ell}(\bx) = A_{\ell}\bx$ and $\bh(\bx)= \sum_{\ell} q_{\ell} \bh_{\ell}(\bx)=A\bx$.


For example, consider $q_0=\alpha$, $q_1=1-\alpha$, $\cA=\{0,1\}$, and $\bx=(x^{00},x^{01},x^{10},x^{11})$. From Theorems~\ref{thm:TDS_SA} and~\ref{thm:k0}, it can be shown that
\begin{align*}
    A_0&=\begin{pmatrix}
    -2 & 1 & 1 & 0\\
    1 & -2 & 0 & 1\\
    1 & 0 & -2 & 1\\
    0 & 1 & 1 & -2
\end{pmatrix},&
A_1&=\begin{pmatrix}
0 & 1 & 0 & 0\\
0 & -1 & 0 & 0\\
0 & 0 & -1 & 0\\
0 & 0 & 1 & 0
\end{pmatrix}.
\end{align*}
and 
\begin{equation}\label{eq:A-ex}
    A=\begin{pmatrix}-2\alpha & 1 & \alpha & 0\\
\alpha & -\left(1+\alpha\right) & 0 & \alpha\\
\alpha & 0 & -\left(1+\alpha\right) & \alpha\\
0 & \alpha & 1 & -2\alpha
\end{pmatrix}.
\end{equation}

\begin{thm}
Consider a tandem duplication and substitution system with distribution $q=\left(q_{\ell}\right)_{0\leq\ell< M}$ over these mutations, with $q_0<1$, and let $A$ be the matrix defined for this system by~\eqref{eq:Adef}. The frequencies of substrings $\bu$ of length $k\ge M$, $\left(x^{\bu}\right)_{\bu\in\cA^k}$, converge almost surely to the null space of the matrix $A$.\label{thm:null}
\end{thm}
\begin{IEEEproof}[Proof]
We first show that the resulting ODE is stable. This is done by applying the Gershgorin circle theorem to the columns of $A$ (see e.g.,~\eqref{eq:A-ex}). In each column, the diagonal element is the only element that can be negative. We show that each column of $A$ sums to 0, which implies that the rightmost point of each circle is the origin. Thus, each eigenvalue of $A$ is either 0 or has a negative real part. 

We show that each column of $A_{\ell}$ sums to zero for each $\ell$, which implies the desired result. Fix $\bv\in U$ and consider the column in $A_{\ell}$ that corresponds to $x^\bv$. To identify the elements in this column, we must consider expressions for $h_{\ell}^\bu(\bx)=\delta_{\ell}^\bu(\bx)-\ell x^{\bu}$ and check if $x^\bv$ appears on the right side. For $\ell>0$, the only negative term corresponds to $h_{\ell}^\bv$, where the coefficient is $-(k-1)$. Inspecting the proofs of Lemmas~\ref{lem:case1}--\ref{lem:case3} shows that for each value of $b\in[k-1]$, there is only one $\bu$ such that $x^\bv$ appears in $h_{\ell}^{\bu}$ with a nonnegative coefficient, and the coefficient is 1. For example, for $b=1$, from Lemma~\ref{lem:case1}, this $\bu$ is equal to $\bv_{\ell}\bv_{1,k-1}$. Since there are $k-1$ possible choices for $b$, the sum of every column in $A_{\ell}$ is 0, as desired. For $\ell=0$, we have $h_{\ell}^\bu(\bx)=\delta_{\ell}^\bu(\bx)$, where $\delta_{\ell}^\bu(\bx)$ is given in Theorem~\ref{thm:k0}. The column corresponding to $x^\bv$ has a negative term equal to $-k$ and $k(|\cA|-1)$ positive terms, where each of the positive terms is equal to $\frac1{|\cA|-1}$, so the sum is again 0.

We have shown that all eigenvalues are either 0 or have negative real parts. For any valid initial point $\bx_0$, the sum of the elements must be 1. Furthermore, each element must be nonnegative. The fact that the columns of $A$ sum to 0 shows that the sum of the elements of any solution $\bx_t$ also equals 1 as $d\bx_t/dt = A\bx_t$. Furthermore, since only diagonal terms in $A$ can be negative, each element of $\bx_t$ is also nonnegative. Thus $\bx_t$ is bounded.

\longversion{
From the Jordan canonical form theorem, we can write $A=PJP^{-1}$ for an invertible matrix of generalized eigenvectors $P$, where $J=\begin{pmatrix}
J' & 0\\
0 & J''
\end{pmatrix}$ and $J'$ and $J''$ are square matrices corresponding to eigenvalue $\lambda=0$ and other eigenvalues respectively. Let $\by_t =P^{-1}\bx_t $, so that $\dot{\by}_t =J\by_t $, which we can write in the form $\dot{\bu}_t =J'\bu_t $ and $\dot{\bw}_t =J''\bw_t $ with $\by_t =(\bu_t ,\bw_t )^T $.
Let $C$ be any compact internally chain transitive set of the ODE
$\dot{\by}_t =J\by_t $. We first show that if $\by=(\bu,\bw)\in C$,
then $\bw=0$. Consider the flow starting from $\by_0 =(\bu_0 ,\bw_0 )^T \in C$
with $\bw_0 \neq\boldsymbol{0}$. We have $\bw_t =e^{J''}\bw_0 $.
Since $J''$ has only eigenvalues with negative real parts,
$\|\bw_t \|\le c_0 e^{-c_1 t}\|\bw_0 \|$ for $t\ge0$ and some
constants $c_0 ,c_1 >0$. If $\by=(\bu,\bw)\in C,$ then $\bw$
is also in an internally chain transitive set of lower dimension.
For $T,\epsilon>0$, let $\bw^{(1)},\dotsc,\bw^{(n)}=\bw^{(1)}$ be
a chain of points such that the flow of $\dot{\bw}_t =J''\bw_t $
starting at $\bw^{(i)}$ meets the $\epsilon$-neighborhood of $\bw^{(i+1)}$
after a time $\ge T$. We thus have 
\begin{equation}
\|\bw^{(i+1)}\|\le c_0 e^{-c_1 T}\|\bw^{(i)}\|+\epsilon.\label{eq:step}\\
\end{equation}
Since $T,\epsilon$ are arbitary, we choose them such that $c_0 e^{-c_1 T}<1/2$
and $c_0 e^{-c_1 T}\|\bw^{(1)}\|<\epsilon<\|\bw^{(1)}\|/2$ if $\|\bw^{(1)}\|>0$.
Hence, $\|\bw^{(2)}\|\le c_0 e^{-c_1 T}\|\bw^{(i)}\|+\epsilon<2\epsilon$
and by induction $\|\bw^{(i+1)}\|\le c_0 e^{-c_1 T}\|\bw^{(i)}\|+\epsilon<2\epsilon$
for $i>1$. This leads to a contraction since it implies that $\|\bw^{(n)}\|=\|\bw^{(1)}\|<2\epsilon$.
Thus $\|\bw^{(1)}\|=0$ and for any $\by=(\bu,\bw)^T \in C$ we must
have $\bw=\boldsymbol{0}$. 

Next, note that since $\bx_t $ is bounded, so is $\by_t $. Hence
for $\by=(\bu,\boldsymbol{0})^T \in C$, $e^{J't}\bu$ must be a
constant since it contains no exponential terms ($\lambda=0$) and
cannot contain a polynomial term in $t$ with degree $\ge1$ (because
of boundedness). So all flows initiated in $C$ are constant. The
same must hold for all flows in $D$, for any $D$ that is an internally
chain transitive invariant set of the ODE $\dot{\bx}_t =A\bx_t $.
Hence, any point in $\bx\in D$ must be in the null space of $A$,
that is, $A\bx=0$. 
}
\end{IEEEproof}

For the matrix $A$ of~\eqref{eq:A-ex}, for $ 0<\alpha<1$, the vector in the null space whose elements sum to 1, and thus the limit of $\bx_n$, is
\begin{equation}\label{eq:binEx}
    \frac1{2(1+3\alpha)}\left({\alpha+1}, {2\alpha},{2\alpha},{\alpha+1}\right)^T.
\end{equation}
If we let $\alpha = \frac{1}{4}$ as an example, the limit of $\bx_n$ then is 
\begin{align}
    \lim\limits_{n\rightarrow \infty}(x_n^{00}, x_n^{01}, x_n^{10}, x_n^{11})^{T} = (\frac{5}{14}, \frac{1}{7}, \frac{1}{7}, \frac{5}{14})^{T}.\label{eq:thresult}
\end{align}

\begin{figure}
    \centering
    \includegraphics[width= 0.5\textwidth]{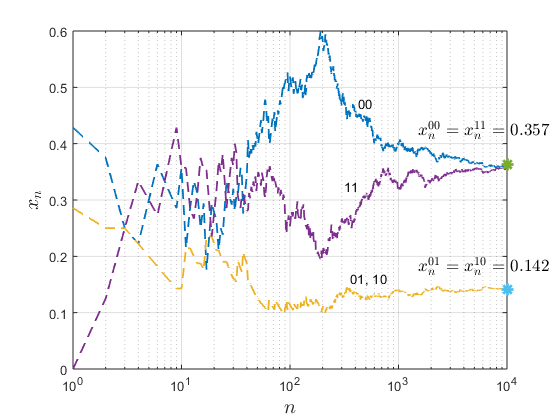}
    \caption{2-mer frequencies vs the number of mutations in a tandem duplication and substitution system, with $\cA = \{0,1\}, \bs_0 = 0100010,q_0 = \frac{1}{4},\text{ and }q_1 = \frac{3}{4}$.}
    \label{fig: sim_tds}
\end{figure}
Figure~\ref{fig: sim_tds} shows the result of simulation of the above TDS system, where $\cA = \{0,1\}$, $\bs_0 = 0100010, q_0 = \frac{1}{4} \text{ and } q_1 = \frac{3}{4}$. As the number $n$ of mutations increases, the frequency vector $\bx_n$ converges to the analytical result (\ref{eq:thresult}). Note that the limits do not depend on the initial sequence $\bs_0$.

Let us consider the two extreme cases. As $\alpha\to1$, all four 2-substrings become equally likely, each with probability $1/4$. Note however that our analysis is not applicable to $q_0=\alpha=1$ since the condition $\sum_n1/|\bs_n|^2 < \infty$ is not satisfied. On the other hand, for a small probability of substitution, $0<\alpha\ll1$, almost all 2-substrings are either $00$ or $11$, as expected. For $\alpha=0$, the null space is spanned by $\bz_1=(1,0,0,0)^T$ and $\bz_2=(0,0,0,1)^T$ and the limit set is $\{a\bz_1+(1-a)\bz_2:0\le a\le 1\}$. In this case, the asymptotic behavior of $k$-mer frequencies will depend on the initial sequence $\bs_0$.

\subsection{Bounds on Entropy\label{subsec: entropy}}
We now turn to provide upper bounds on the entropy. 
We first formally define the entropy, and then argue that the entropy is upper bounded by the capacity of an appropriately defined semiconstrained system~\cite{elishco2016semiconstrained, elishco2017encoding, elishco2018independence}.

Consider the string $\bs_n$, obtained from $\bs_0$ by $n$ rounds of mutations, as described previously. Its expected length is $\ev[\abs{\bs_n}]=\abs{\bs_0}+n\sum_{\ell=1}^{M-1} \ell q_{\ell}$.
We define the entropy after $n$ rounds as
\begin{equation}
    \cH_n=-\frac{1}{\ev[\abs{\bs_n}]}\sum_{\bw\in\cA^*}\Pr(\bs_n=\bw)\log_{\abs{\cA}}\Pr(\bs_n=\bw),\label{eq:entropy}
\end{equation}
and the entropy $\cH_\infty=\limsup_{n\to\infty}\cH_n$.

It is common to define the entropy of DNA sequences based on the limit of block entropies~\cite{lio1996,schmitt1997,herzel1994entropies}. Specifically, let $h_k=-\sum_{\bu\in{\Sigma^k}}p_\bu \log p_\bu$, where $p_\bu$ is the probability of observing $\bu$. Entropy is then obtained as $h_{k+1}-h_k$ for $k\to\infty$. This definition may lead to misleading results. For example, consider a string system in which $\bs_n$ is the De Bruijn sequence of order $n$ (which contains all strings of length $n$ precisely once), obtained according to some deterministic algorithm. Based on block entropies, the entropy of the system can be shown to equal $\log |\Sigma|$, while the system is in fact deterministic. The definition in~\eqref{eq:entropy} gives the correct entropy, i.e., $0$, since there is only one possibility for $\bs_n$ for each $n$.

Let us recall some definitions concerning semiconstrained systems (see \cite{elishco2018independence}). Fix $k$ and let $\cP(\cA^k)$ denote the set of all probability measures on $\cA^k$. A \emph{semiconstrained system} is defined by $\Gamma_k\subseteq\cP(\cA^k)$. The set of the admissible words of the semiconstrained system, denoted $\cB(\Gamma_k)$, contains exactly all finite words over the alphabet $\cA$ whose $k$-mer distribution is in $\Gamma_k$. Let $\cB_n(\Gamma_k)=\cB(\Gamma_k)\cap\cA^n$. An expansion of $\Gamma_k$ by $\epsilon>0$ is defined as
\[ \bbB_\epsilon(\Gamma_k) = \mathset{ \bxi\in\cP(\cA^k) ~:~ \inf_{\bnu\in\Gamma_k}\|\bnu-\bxi\|_{\mathrm{TV}}\leq \epsilon}, \]
where $\|\cdot\|_{\mathrm{TV}}$ denotes the total-variation norm. \longversion{Thus, $\bbB_\epsilon(\Gamma_k)$ contains all the measures in $\Gamma_k$ as well as those which are $\epsilon$-close to some measure in $\Gamma_k$.} The capacity of $\Gamma_k$ is then defined as
\[\ccap(\Gamma_k)=\lim_{\epsilon\to 0^+}\limsup_{n\to\infty}\frac{1}{n}\log_{\abs{\cA}}\abs{\cB_n(\bbB_\epsilon(\Gamma_k))},\]
which intuitively measures the information per symbol in strings whose $k$-mer distribution is in (or ``almost'' in) $\Gamma_k$.

\begin{thm}
For the mutation process described above, for $k\in\bbN^+$, if the vector of the frequencies $\bx$ of strings of length $k$ converges almost surely to a set $\Gamma_k$, then $\cH_\infty\leq \ccap(\Gamma_k)$ \label{thm:bounds on entropy}.
\end{thm}
\longversion{
\begin{IEEEproof}
Fix some positive real number $\epsilon>0$. by Hoeffding's inequality,
\[\Pr\parenv{\abs{\abs{\bs_n}-\ev[\abs{\bs_n}]}\geq \epsilon n}\leq 2\exp\parenv{-\frac{2\epsilon^2}{(M-1)^2}n}.   \]
We also note that $\abs{\bs_n}\leq \abs{\bs_0}+(M-1)n$ for all $n$.

We know that $\bx_n$ converges almost surely to some point in $\Gamma_k$ as
$n\to\infty$, and thus, for every $\epsilon >0$, there exists $N(\epsilon)$ such that $\bx_n\in\bbB_\epsilon(\Gamma_k)$ for all $n\geq N(\epsilon)$ with probability at least $1-\epsilon$. Hence, for all large enough $n$,
\begin{align}
&\cH_n = -\frac{1}{\ev(\abs{\bs_n})} \!\!\!\!\sum_{\substack{\bw\in \cA^*\\ \abs{\bw}\leq \abs{\bs_0}+(M-1)n}}\!\!\!\!\!\!\!\Pr(\bs_n=\bw)\log_{\abs{\cA}}\Pr(\bs_n=\bw)\nonumber\\
&\leq \frac{1\!-\!\epsilon\!-\!2\exp\parenv{-\frac{2\epsilon^2}{(M-1)^2}n}}{\ev(\abs{\bs_n})}\log_{\abs{\cA}}\parenv{\sum_{i=\ev(\abs{\bs_n})-\epsilon n}^{\ev(\abs{\bs_n})+\epsilon n}\!\!\!\!\!\!\!\abs{\cB_i(\bbB_\epsilon(\Gamma_k))}}\nonumber\\
&\quad\ + \frac{\epsilon+2\exp\parenv{-\frac{2\epsilon^2}{(M-1)^2}n}}{\ev(\abs{\bs_n})}\log_{\abs{\cA}}\abs{\bigcup_{i=1}^{\abs{\bs_0}+(M-1)n}\cA^i}\nonumber\\
&\quad\ +H_2\parenv{\epsilon+2\exp\parenv{-\frac{2\epsilon^2}{(M-1)^2}n}}\log_{\abs{\cA}}2, \label{eq:mainneq}
\end{align}
where $H_2(x)=-x\log_2 x-(1-x)\log_2 (1-x)$ is the binary entropy function.

Again, by the definition of the capacity of semiconstrained systems, for all large enough $n$,
\[\abs{\cB_i(\bbB_\epsilon(\Gamma_k))}\leq \abs{\Sigma}^{i\cdot\ccap(\bbB_\epsilon(\Gamma_k))+\epsilon}.\]
Plugging this back into the main inequality~(\ref{eq:mainneq}), and further simplifying gives
\begin{align*}
\cH_n &\leq \frac{1}{\ev(\abs{\bs_n})}\log_{\abs{\cA}}\parenv{2\epsilon n\abs{\cA}^{(\ev(\abs{\bs_n})+\epsilon n)(\ccap(\bbB_\epsilon(\Gamma_k))+\epsilon)}}\\
&\quad\ + \frac{\epsilon+2\exp\parenv{-\frac{2\epsilon^2}{(M-1)^2}n}}{\ev(\abs{\bs_n})}(\abs{\bs_0}+(M-1)n+1)\\
&\quad\ +H_2\parenv{\epsilon+2\exp\parenv{-\frac{2\epsilon^2}{(M-1)^2}n}}\log_{\abs{\cA}}2.
\end{align*}
Taking $\limsup_{n\to\infty}$ of both sides we obtain
\begin{align*}
\cH_{\infty}&\leq \parenv{1+\frac{\epsilon}{\sum_{i=1}^{M-1} iq_i}}\cdot\parenv{\ccap(\bbB_\epsilon(\Gamma_k))+\epsilon}\\
&\quad\ +\frac{\epsilon (M-1)}{\sum_{i=1}^{M-1} iq_i}
+H_2(\epsilon)\log_{\abs{\Sigma}}2.
\end{align*}
Finally, taking $\lim_{\epsilon\to 0^+}$ of both sides, we obtain the claim.
\end{IEEEproof}} 

\begin{rem}
We comment that if $\Gamma_k=\mathset{\bxi_k}$, i.e., $\Gamma_k$ contains a single shift-invariant measure\footnote{A shift-invariant measure $\bxi_k\in\cP(\cA^k)$ is a measure that satisfies $\sum_{a\in\cA}\bxi_k^{a\bw}=\sum_{a\in\cA}\bxi^{\bw a}_k$ for all $\bw\in\cA^{k-1}$. The $k$-mer distributions of cyclic strings are always shift invariant, and thus a converging sequence of such measures also converges to a shift-invariant measure.}, then $\ccap(\Gamma_k)$ has a nice form for all $k\in\bbN^{+}$ (see \cite{elishco2018independence,elishco2016semiconstrained}):
\[\ccap(\Gamma_k)=-\sum_{a_1\dots a_k\in\cA^k}\bxi_k^{a_1\dots a_k}\log_{\abs{\cA}}\frac{\bxi_k^{a_1\dots a_k}}{\bar{\bxi}_k^{a_1\dots a_{k-1}}},\]
where $\bar{\bxi}_k$ is the marginal of $\bxi_k$ on the first $k-1$ coordinates, i.e., $\bar{\bxi}_k^{a_1\dots a_{k-1}}=\sum_{b\in\cA}\bxi_k^{a_1\dots a_{k-1}b}$. Furthermore, 
$\forall k\in\bbN^{+}$,
\[\ccap(\Gamma_k)\geq\ccap(\Gamma_{k+1}),\]
which follows from the fact that $\ccap(\Gamma_k)$ can be viewed as the conditional entropy of a symbol given the $k-1$ previous symbols in a stationary process.
\end{rem}

Using the preceding remark and Theorem~\ref{thm:bounds on entropy}, we can find a series of upper bounds on a given system:
\[\ccap(\Gamma_1) \geq \ccap(\Gamma_2) \geq \cdots \ge \ccap(\Gamma_k) \geq \cdots \geq \cH_{\infty},\]
with $\Gamma_k$ being the limit of $(x^{\bu})_{\bu\in\cA^k}$
\begin{figure}
    \centering
    \includegraphics[width = 0.5\textwidth]{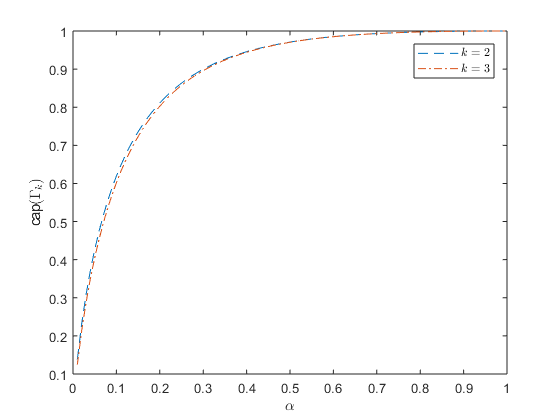}
    \caption{Entropy bound vs the probability of substitution, with $\cA = \{0,1\}$.}
    \label{fig:entbd}
\end{figure}
\begin{figure}
    \centering
    \includegraphics[width = 0.5\textwidth]{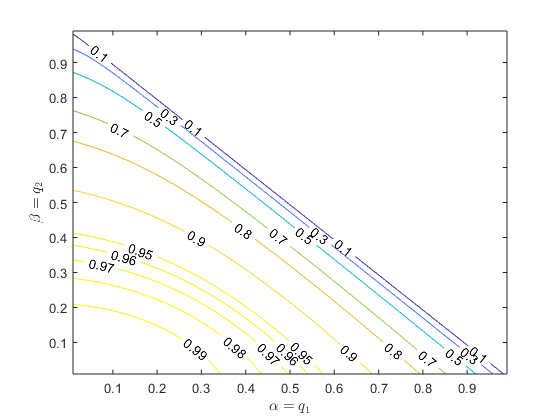}
    \caption{Contour plot of entropy bounds, with $\cA = \{0,1\}$, $k = 3$, $q_0=1-\alpha-\beta$, $q_1 = \alpha$, $q_2 = \beta$. }
    \label{fig:conentbd}
\end{figure}

In particular, for the system whose limit is given by~\eqref{eq:binEx}, we have ${\bxi}^0={\bxi}^1=\nicefrac{1}{2}, \bxi^{00} = \bxi^{11} = \nicefrac{(\alpha + 1)}{2(1+3\alpha)}, \bxi^{01} = \bxi^{10} = \nicefrac{\alpha}{7}$. It then follows that for this system $\cH_\infty\le H_2\left(\frac{2\alpha}{1+3\alpha}\right) = \ccap(\Gamma_2)$. We can also compute $\ccap(\Gamma_k)$ for $k = 3,4,\ldots$. Figure~\ref{fig:entbd} shows the entropy bound we find using 2-mer and 3-mer frequencies. The two bounds are close, which suggests that we may be close to the exact entropy values. However, in the absence of a lower bound, this conjecture cannot be verified. The figure shows that when there is only one possible duplication length, the source of diversity is substitution, as may be expected. As $\alpha \rightarrow 1$, the relative number of substitutions increases, causing $\Gamma_k$ to be close to the uniform distribution, and the entropy tends to 1. On the other hand, as $\alpha \rightarrow 0$, only duplications occur. This leads to the generation of low complexity sequences that consists of long runs of 0s and 1s, and thus entropy that is close to 0. 

Figure~\ref{fig:conentbd} shows the entropy bound computed using 3-mer frequencies for the case in which $\cA=\{0,1\}$, $q_1 = \alpha,q_2 = \beta$ and $q_0 = 1-\alpha-\beta$. So in this system, duplications of lengths 1 and 2 are both possible. It can be seen that similar to Figure~\ref{fig:entbd}, even a small probability of substitution leads to relatively high values of entropy. Furthermore, we note that, as may be expected, longer duplications lead to a smaller value of entropy.
 
\section{Interspersed Duplication\label{sec: ID}}
In this section, we study the evolution of $k$-mer frequencies of the interspersed-duplication system, also using the stochastic-approximation technique. 

Let $U=\bigcup_{i=1}^k \cA^i$, i.e., the set consisting of all non-empty strings of length at most $k$. Also, let the vectors $\xmul n$ and $\mul n$ be defined as before using $U$. 
\begin{thm}
Consider $\bu\in U$.
In an interspersed-duplication system, for $\ell<\left|\bu\right|$, we have 
\begin{align*}
\diff{\ell}\bu=&-\left(\left|\bu\right|-1\right)\xmult n\bu+\sum_{i=1}^{\ell}\xmult n{\bu_{1,i}}\xmult n{\bu_{i+1,\left|\bu\right|-i}}\\
&+\sum_{i=1}^{\ell}\xmult n{\bu_{1,\left|\bu\right|-i}}\xmult n{\bu_{\left|\bu\right|-i+1,i}}+\sum_{i=1}^{\mathclap{\left|\bu\right|-\ell-1}}\xmult n{\bu_{1,i}\bu_{i+\ell+1,\left|\bu\right|-\ell-i}}\xmult n{\bu_{i+1,\ell}}.
\end{align*}
\end{thm}
\begin{IEEEproof}
The term $-\left(\left|\bu\right|-1\right)\xmult n\bu$ accounts
for the expected number of lost occurrences of $\bu$ in $\bs_n$ as a result
of inserting the duplicate substring. To illustrate, assume $\ab=\left\{ \ga,\gc,\gg,\gt\right\} $, $\bu=\ga\gc\gt$ and $\ell=1$. An occurrence of
$\bu=\ga\gc\gt$ will be lost if for example an occurrence of the symbol $\gg$ is duplicated and
inserted after $\ga$ in this occurrence of $\bu$, since it becomes
$\ga\gg\gc\gt$. The probability that a certain occurrence is lost
equals $\frac{\left|\bu\right|-1}{L_{n}}$. Since there are $\mult n\bu$
such occurrences, the expected number of lost occurrences of $\bu$
equals $\mult n\bu\frac{\left|\bu\right|-1}{L_{n}}=\xmult n\bu\left(\left|\bu\right|-1\right)$.
Note that if the symbol $\gt$ is duplicated and inserted after $\gc$
in an occurrence of $\ga\gc\gt$, we still count the original occurrence
as lost, but count a new occurrence in the resulting $\ga\gc\gt\gt$,
as seen in what follows. We now explain the first summation above.
This summation represents the newly created occurrences of $\bu$ where
the first $i$ symbols come from the duplicate and the next $\left|\bu\right|-i$
are from the substring that starts after the point of insertion of
the duplicate. There are $\mult n{\bu_{1,i}}$ occurrences of $\bu_{1,i}$.
The duplicate ends with one of these with probability $\frac{\mult n{\bu_{1,i}}}{L_{n}}=\xmult n{\bu_{1,i}}$.
Furthermore, the duplicate is inserted before an occurrence of $u_{i+1,\left|\bu\right|-i}$
with probability $\xmult n{\bu_{i+1,\left|\bu\right|-i}}$. Hence, the
probability of a new occurrence created in this way is $\xmult n{\bu_{1,i}}\xmult n{\bu_{i+1,\left|\bu\right|-i}}$,
and so is the expected number of such new occurrences. The role of
the second summation is similar, except that the duplicate provides
the second part of $\bu$. The last summation accounts for new occurrences
of $\bu$ in which the duplicate substring forms a middle part of $\bu$
of length $\ell$ and previously existing substrings contribute a
prefix of length $i$ and a suffix of length $\left|\bu\right|-\ell-i$.
In terms of our running example with $\bu=\ga\gc\gt$ and $\ell=1$,
one such new occurrence is created if $\gc$ is duplicated and inserted
after $\ga$ in an occurrence of $\ga\gt$. The probability of such
an event is $\xmult n{\bu_{1,i}\bu_{i+\ell+1,\left|\bu\right|-\ell-i}}\xmult n{\bu_{i+1,\ell}}=\xmult n{\ga\gt}\xmult n{\gc}$,
where $i=1$.
\end{IEEEproof}
\begin{thm}
For $\ell\ge\left|\bu\right|$, we have 
\begin{align*}
\diff{\ell}\bu & =-\left(\left|\bu\right|-1\right)\xmult n\bu+\sum_{i=1}^{\left|\bu\right|-1}\xmult n{\bu_{1,\left|\bu\right|-i}}\xmult n{\bu_{\left|\bu\right|-i+1,i}}\\
 & \quad+\sum_{i=1}^{\left|\bu\right|-1}\xmult n{\bu_{1,i}}\xmult n{\bu_{i+1,\left|\bu\right|-i}}+\left(\ell-\left|\bu\right|+1\right)\xmult n\bu\\
\end{align*}
\end{thm}
\begin{IEEEproof}
The first two summations are similar to the first two summations
for the case of $\ell<\left|\bu\right|$, but a term corresponding to
the third summation is not present. The term $\left(\ell-\left|\bu\right|+1\right)\xmult n\bu$
corresponds to the cases in which a new occurrence of $\bu$ is created
as a substring of the duplicate substring. 
\end{IEEEproof}
Note that $\diff{\ell}\bu$
depends only on $\xmul n$ and is Lipschitz since $\xmul n\in\left[0,1\right]^{\left|U\right|}$. Thus, (A.\ref{asm:delta-f-of-x}) and (A.\ref{asm:delta-lipshitz}) hold.

Since $\hmult{\ell}\bu\left(\bx_n\right)=\diff{\ell}\bu\left(\bx_n\right)-\ell x_n^{\bu}$
, we have for $\ell<\left|\bu\right|$ and $\ell\ge\left|\bu\right|$, respectively,
\begin{align}
h_{\ell}^{\bu}(\bx_n) &= -(\ell + \abs{\bu} -1)x_{n}^{\bu} + \sum\limits_{i = 1}^{\ell}x_n^{\bu_{1,i}}x_{n}^{\bu_{i+1,\abs{\bu}-i}} \label{eq:randCric_k<l} \\
+\!\!&\sum\limits_{i = 1}^{\ell}\! x_{n}^{\bu_{1,\abs{\bu}-i}}x_{n}^{\bu_{\abs{\bu}-i+1,i}} 
\!\!+\!\!\!\!\!\!\!\sum\limits_{i = 1}^{\abs{\bu}-\ell-1}\!\!\!\!\!\!x_{n}^{\bu_{1,i}\bu_{i+\ell+1,\abs{\bu}-\ell-i}}x_{n}^{\bu_{i+1,\ell}}\nonumber \\ 
h_{\ell}^{\bu}(\bx_{n})&= -2(\abs{\bu}-1)x_{n}^{\bu} + 2\sum\limits_{i = 1}^{\abs{\bu}-1}x_{n}^{\bu_{1,i}}x_{n}^{\bu_{i+1,\abs{\bu}-i}}
 \label{eq:randCirc_k>=00003Dl}
\end{align}
Recall that $\hmul{\ell}\left(\xmul{}\right)=\left(h_{\ell}^{\bu}\left(\xmul{}\right)\right)_{\bu\in U}$.
So from~(\ref{eq:randCric_k<l}) and~(\ref{eq:randCirc_k>=00003Dl}),
we can find the ODE $d\xmul t/dt=\hmul{}\left(\xmul t\right)=\sum_{\ell=1}^{M-1}q_{\ell}\hmul{\ell}\left(\xmul t\right)$. As an example, if $k=3$ and $\mathcal{A}=\left\{ \ga,\gc\right\} $,
then 
$
U=(\ga,\gc,\ga\ga,\ga\gc,\gc\ga,\gc\gc,\ga\ga\ga,\dotsc,\gc\gc\gc)
$
 and some of the equations of the ODE system are
\begin{align}
\frac{d}{dt}\xmult t{\ga\phantom{\ga\gc}} & =\frac{d}{dt}\xmult t{\gc}=0,\nonumber \\
\frac{d}{dt}\xmult t{\ga\ga\phantom{\gc}} & =-2\xmult t{\ga\ga}+2\left(\xmult t{\ga}\right)^{2},\quad\frac{d}{dt}\xmult t{\ga\gc}=-2\xmult t{\ga\gc}+2\xmult t{\ga}\xmult t{\gc},\nonumber \\
\frac{d}{dt}\xmult t{\ga\ga\gc} & =-\left(4-q_{1}\right)\xmult t{\ga\ga\gc}+2\xmult t{\ga}\xmult t{\ga\gc}+\left(2-q_{1}\right)\xmult t{\gc}\xmult t{\ga\ga}.\label{eq:diff2}
\end{align}

For a vector $\xmul{}$ that contains the elements $\left(\xmult{}a\right)_{a\in\mathcal{A}}$
and for $\bv\in\mathcal{A}^{*}$, define 
\(
p\left(\bv,\xmul{}\right)=\prod_{a\in\ab}\left(\xmult{}a\right)^{n_{\bv}\left(a\right)}
\), where $n_{\bv}(a)$ is the number of occurrences of $a$ in $\bv$, 
and note that $p\left(\bv\bw,\xmul{}\right)=p\left(\bv,\xmul{}\right)p\left(\bw,\xmul{}\right)$.
We now turn to find the solutions to the ODE $d\xmul t/dt=\hmul{}\left(\xmul t\right)$.
\begin{lem}
\label{lem:sol-inter}Consider the ODE $d\xmul t/dt=\hmul{}\left(\xmul t\right)$
where $\hmul{}\left(\xmul{}\right)=\sum_{\ell=1}^{M-1}q_{\ell}\hmul{\ell}\left(\xmul{}\right)$
and the elements of $\hmul{\ell}\left(\xmul{}\right)$ are given by
(\ref{eq:randCric_k<l}) and (\ref{eq:randCirc_k>=00003Dl}). The
solution to this ODE is 
\begin{equation}
\xmult t\bv=p\left(\bv,\xmul 0\right)+\sum_{i}b_{i}^{\bv}e^{-d_{i}^{\bv}t},\qquad \bv\in U,\label{eq:odesol-inter}
\end{equation}
where $\xmul 0=\left.\xmul t\right|_{t=0}$; the range of $i$ in
the summation is finite; and $b_{i}^{\bv}$ and $d_{i}^{\bv}$ are constants
with $d_{i}^{\bv}>0$.\end{lem}
\begin{IEEEproof}
We prove the lemma by induction. The claim (\ref{eq:odesol-inter})
holds for $\bv\in\mathcal{A}$, since the equations for $\xmult ta$,
$a\in\mathcal{A}$, are of the form $d\xmult ta/dt=0$ and so $\xmult ta=\xmult 0a$.
Fix $\bu\in U$ such that $\left|\bu\right|>1$, and assume that (\ref{eq:odesol-inter})
holds for all $\bv\in U$ such that $\left|\bv\right|<\left|\bu\right|$.
We show that it also holds for $\bu$, i.e., 
\(
\xmult t\bu=p\left(\bu,\xmul 0\right)+\sum_{i}b_{i}^{\bu}e^{-d_{i}^{\bu}t}.
\)
Using the assumption, we rewrite (\ref{eq:randCric_k<l}) and (\ref{eq:randCirc_k>=00003Dl})
as 
\begin{align*}
\hmult{\ell}\bu(\xmul t) & =-\left(\ell+\left|\bu\right|-1\right)\left(\xmult t\bu-p\left(\bu,\xmul 0\right)\right)+\sum_{i}b_{i}^{'}e^{-d_{i}^{'}t}
\end{align*}
for $\ell<\left|\bu\right|$, and 
\[
\hmult{\ell}\bu\left(\xmul t\right)=-2\left(\left|\bu\right|-1\right)\left(\xmult t\bu-p\left(\bu,\xmul 0\right)\right)+\sum_{i}b_{i}^{''}e^{-d_{i}^{''}t}
\]
for $\ell\ge\left|\bu\right|$, where $b_{i}^{'},d_{i}^{'},b_{i}^{''},d_{i}^{''}$
are constants with $d_{i}^{'},d_{i}^{''}>0$. Hence, $\hmult{}\bu\left(\xmul t\right)$
can be written as 
\[
\hmult{}\bu\left(\xmul t\right)=-c^{\bu}\left(\xmult t\bu-p\left(\bu,\xmul 0\right)\right)+\sum_{i}b_{i}^{'''}e^{-d_{i}^{'''}t},
\]
where $c^{\bu}=2\left|\bu\right|-2-\sum_{\ell=1}^{\left|\bu\right|-1}q_{\ell}\left(\left|\bu\right|-1-\ell\right)$,
and $b_{i}^{'''},d_{i}^{'''}$ are constants with $d_{i}^{'''}>0$.
Thus the solution to the ODE $d\xmult t\bu/dt=\hmult{}\bu\left(\xmul t\right)$
is
\begin{align*}
\xmult t\bu & =e^{-c^{\bu}t}\!\!\int\! e^{c^{\bu}t^{'}}\!\!\Bigl(c^{\bu}p\left(\bu,\xmul 0\right)+\!\sum_{i}b_{i}^{'''}e^{-d_{i}^{'''}t^{'}}\Bigr)dt^{'}\!+\bar{b}e^{-c^{\bu}t}\\
 & =p\left(\bu,\xmul 0\right)+\sum_{i}b_{i}^{\bu}e^{-d_{i}^{\bu}t},
\end{align*}
where $\bar{b},b_{i}^{\bu},d_{i}^{\bu}$ are some constants, with $d_{i}^{\bu}>0$
(note that $c^{\bu}>0$ since $\left|\bu\right|>1$). This completes the
proof.
\end{IEEEproof}
For example, the solutions to (\ref{eq:diff2}) with $q_{1}=0$ are
\begin{align*}
\xmult t{\ga\phantom{\ga\gc}} & =\xmult 0{\ga},\qquad\xmult t{\gc}=\xmult 0{\gc},\\
\xmult t{\ga\ga\phantom{\gc}} & =\left(\xmult 0{\ga}\right)^{2}+b_{1}^{\ga\ga}e^{-2t},\quad\xmult t{\ga\gc}=\xmult 0{\ga}\xmult 0{\gc}+b_{1}^{\ga\gc}e^{-2t},\\
\xmult t{\ga\ga\gc} & =\left(\xmult 0{\ga}\right)^{2}\xmult 0{\gc}+b_{1}^{\ga\ga\gc}e^{-2t}+b_{2}^{\ga\ga\gc}e^{-4t},
\end{align*}
where $b_{1}^{\ga\ga\gc}=\xmult 0{\ga}b_{1}^{\ga\gc}+\xmult 0{\gc}b_{1}^{\ga\ga}$.

In the next theorem, we use Lemma \ref{lem:sol-inter} to characterize
the limits of the frequencies of substrings in interspersed-duplication
systems.
\begin{thm}\label{thm:ID}
Let $U=\bigcup_{i=1}^k \cA^i$, and let $\xmul
n=\left(\xmult n\bu\right)_{\bu\in U}$ be the vector of frequencies of
these strings at time $n$ in an interspersed-duplication system. The
vector $\xmul n$ converges almost surely.  Furthermore, its limit
$\xmul{\infty}$ satisfies
\[
\xmult{\infty}\bu=\prod_{a\in\ab}\left(\xmult{\infty}a\right)^{n_{\bu}\left(a\right)},\quad\mbox{for all }\bu\in U.
\]
\end{thm}
Note that the existence of the limits $\xmult \infty a$ of $\xmult na$, for $a\in\mathcal A$, was also shown in Theorem~\ref{thm:singleton1}.
\begin{IEEEproof}
From Theorem \ref{thm:borkar}, we know that the limit set of $\xmul n$
is an internally chain transitive invariant set of the ODE described
by (\ref{eq:randCric_k<l}) and (\ref{eq:randCirc_k>=00003Dl}). Let
this set, which consists of points of the form $\ymul{}=\left(\ymult{}\bv\right)_{\bv\in U}$,
be denoted by $H$. Since for each $\bu\in U$, $\xmult n\bu\in\left[0,1\right]$,
we can assume that $H\subseteq\left[0,1\right]^{\left|U\right|}$ without any loss of generality. We now
use these facts to show that $\ymult{}\bu=p\left(\bu,\ymul{}\right)$ for each $\ymul{}\in H$ and $\bu\in U$.

Suppose to the contrary that there exist $\ymul{}\in H$ and $\bu\in U$
such that $\ymult{}\bu\neq p\left(\bu,\ymul{}\right)$. Among all possible
choices for such $\ymul{}$ and $\bu$, choose the ones where the length
$\left|\bu\right|$ of $\bu$ is minimum. Hence, $\ymult{}\bu\neq p\left(\bu,\ymul{}\right)$
but $\zmult{}\bv=p\left(\bv,\zmul{}\right)$ for all $\bv\in\mathcal{A}^{*}$
with $\left|\bv\right|<\left|\bu\right|$, and all $\zmul{}\in A$. Then, similar to the proof of Lemma \ref{lem:sol-inter}, one can show that
if $\xmul 0=\zmul{}\in A$, then $\xmult t\bu=p\left(\bu,\zmul{}\right)+be^{-c^{\bu}t},$
where $b=\zmult{}\bu-p\left(\bu,\zmul{}\right)$ and $c^{\bu}\ge\left|\bu\right|$.

By the definition of internal chain transitivity, for any $\epsilon>0$
and $T>0$, there exist $N\ge1$ and a sequence $\ymul 0,\dotsc,\ymul N$
with $\ymul i\in H$, $\ymul 0=\ymul N=\ymul{}$ such that for $0\le i<N$,
if $\xmul 0=\ymul i$, then there exists $t\ge T$ such that $\xmul t$
is in the $\epsilon$-neighborhood of $\ymul{i+1}$. Suppose $\xmul 0=\ymul i$
and suppose for $t'\ge T$, $\xmul {t'}$ is in the $\epsilon$-neighborhood
of $\ymul{i+1}$. We have
\begin{align}
\abs{\ymult{i+1}\bu \!-\!\xmult{t'}\bu} = \abs{\ymult{i+1}\bu\! - \!p\left(\bu,\ymul i\right)\!- \!\left(\ymult i\bu\!-\!p\left(\bu,\ymul i\right)\right)e^{-c^{\bu}t'} }      \leq \epsilon. \label{eq: general flow inequation}
\end{align}
Furthermore, since for $a \in \cA$, $x_{t'}^a = p(a, \by_i) = y_i^a$, we also have 
\begin{align}
    \abs{y_{i+1}^{a} - y_i^a} \leq \epsilon \label{eq: ya}
\end{align}
So if $p(\bu,\by_{i+1})>0$, we have, 
\begin{align}
&p(\bu,\by_{i}) - p(\bu,\by_{i+1}) \nonumber\\
 \leq&  \prod_{a\in\cA}(y_{i+1}^a)^{n_{\bu}(a)} \left(\prod_{a\in\cA}\left(\frac{y_{i+1}^a + \epsilon}{y_{i+1}^a}\right)^{n_{\bu}(a)} -1\right)\nonumber\\
 \leq & \prod_{a\in\cA}\left(1+\frac{\epsilon}{y_{i+1}^{a}}\right)^{n_{\bu}(a)} -1\nonumber\\
 \leq & \left(1+\frac{\epsilon}{\min\limits_{a\in\cA}y_{i+1}^a}\right)^{\abs{\bu}} -1, \label{eq:p>0}
\end{align}
and if $p(\bu,\by_{i+1}) =0$, 
\begin{align}
    p(\bu,\by_{i}) - p(\bu,\by_{i+1}) \leq \epsilon^{n_{\bu}(a)}. \label{eq:p=0}
\end{align}\\
Thus, from (\ref{eq: general flow inequation}), (\ref{eq:p>0}) and (\ref{eq:p=0}), it follows that 
\begin{align}
\ymult{i+1}\bu-p\left(\bu,\ymul{i+1}\right)\le e^{-c^{\bu}T}+\epsilon + O(\epsilon).\label{eq:y-psmall}
\end{align}
In particular, (\ref{eq:y-psmall}) holds for $i=N-1$, i.e.,
\[y^{\bu} - p(\bu,\by) \leq e^{-c^{\bu}T} + O(\epsilon).\]
But we can make the right side of the above inequalities arbitrary small
by choosing $T$ large enough and $\epsilon$ small enough. Thus $\ymult{}\bu=p\left(\bu,\ymul{}\right)$,
which is a contradiction. Hence, for each $\ymul{}\in H$ and $\bu\in U$,
we have $\ymult{}\bu=p\left(\bu,\ymul{}\right)$, and the theorem follows. 
\end{IEEEproof}

\begin{figure}
\includegraphics[trim = 5mm 4mm 7mm 6mm, clip, width=1\columnwidth]{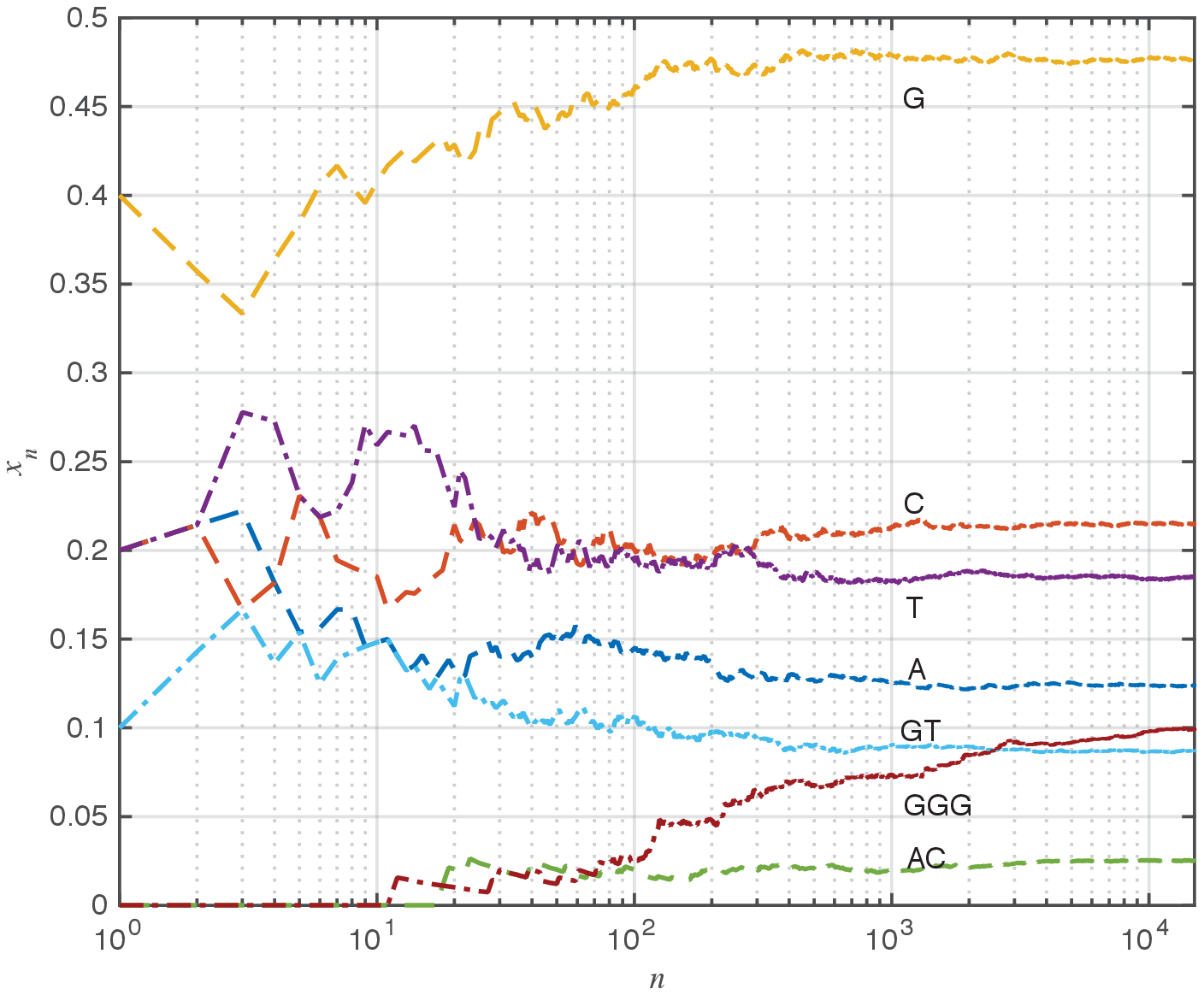}\caption{\label{fig:inters}Symbol frequencies vs the number of duplications
in an interspersed-duplication system, with $\bs_0=\ga\gg\protect\gc\protect\gg\protect\gt\protect\ga\protect\gt\protect\gg\protect\gc\protect\gg$,
and $q_{4}=q_{6}=\nicefrac{1}{2}$.}
\end{figure}


The theorem shows that for $\bu\in\mathcal{A}^{*}$, the frequency
of $\bu$ converges to the frequency of same in an iid sequence where
the probability of $a\in\mathcal{A}$ equals $\xmult{\infty}a$. Figure
\ref{fig:inters} illustrates an example, obtained via simulation,
where the system starts with $\bs_0=\ga\gg\gc\gg\gt\ga\gt\gg\gc\gg$
and duplications of lengths 4 and 6 occur with equal probability.
As the number $n$ of duplications increases, the frequency vector
$\xmul n$ becomes more compatible with that of an iid sequence. For
example for $n=15000$, we have 
$\xmult n{\ga\gc}  =0.0251\simeq\xmult n{\ga}\xmult n{\gc}=0.0266,$
$\xmult n{\gg\gt}  =0.0872\simeq\xmult n{\gg}\xmult n{\gt}=0.0880,$ and
$\xmult n{\gg\gg\gg}  =0.0992\simeq\left(\xmult n{\gg}\right)^{3}=0.1084$

The limit set for $(x^\bu)_{\bu\in\cA^k}$ implied by Theorem~\ref{thm:ID} includes the uniform distribution. As a result, the application of Theorem~\ref{thm:bounds on entropy} leads to the trivial upper bound of $|\cA|$. It thus appears that determining the entropy of ID systems requires determining not only the limit set for the $k$-mer frequencies but also their limiting distribution, as well as results that can relate this distribution to entropy. We leave pursuing this direction to future work. Nevertheless, the fact that the $k$-mer frequencies are similar to those in iid sequences, suggest that interspersed duplication leads to high entropy. At least for certain special cases, this is indeed the case. For binary ($\cA=\{0,1\}$) interspersed duplications of length 1, the entropy is found in~\cite{elishco2018} as 
\[
\frac{\log_2e}{t_0+t_1}((t_0+t_1)\mathsf{H}_{t_0+t_1}-t_0\mathsf{H}_{t_0}-t_1\mathsf{H}_{t_1}),
\]
where $t_0$ and $t_1$ are the numbers of $0$s and $1$s in $\bs_0$, respectively, and $\mathsf{H_t}$ is the $t$th Harmonic number. For $t_0=t_1\to\infty$, the entropy can be shown to equal 1.

\section{Conclusion\label{sec: Con}}

We studied the limiting behavior of two stochastic duplication systems, tandem duplication with substitution and interspersed duplication. We used stochastic approximation to compute the limits of $k$-mer frequencies for tandem duplications and substitutions. We also provided a method for determining upper bounds on the entropy of these systems. For interspersed duplication system, we 
established that $k$-mer frequencies tend to the corresponding probabilities in sequences generated
by iid sources. This suggests that these systems have high entropy, and the structure of the limit set for $k$-mers prevents us from obtaining non-trivial upper bounds. Many problems are left open. First, for tandem duplication and substitution systems, other mutations, such as deletions were not studied; and for interspersed duplication, substitutions, deletions and other mutations were not considered. Moreover, for interspersed duplication, providing nontrivial upper bounds on the entropy requires further research. While we conjecture that the upper bounds presented here are close to actual values, lower bounds on the entropy are needed to verify this claim. Since this work was limited
to the asymptotic analysis of these systems, more research is required
to quantify their finite-time behavior. 

\printbibliography

\end{document}